%% file: sn-article-arxiv.tex
\documentclass[pdflatex,sn-basic]{sn-jnl}

\usepackage{graphicx}
\usepackage{multirow}
\usepackage{amsmath,amssymb,amsfonts}
\usepackage{amsthm}
\usepackage{mathrsfs}
\usepackage[title]{appendix}
\usepackage[table]{xcolor}
\usepackage{textcomp}
\usepackage{manyfoot}
\usepackage{booktabs}
\usepackage{algorithm}
\usepackage{algorithmicx}
\usepackage{algpseudocode}
\usepackage{listings}
\usepackage{float}
\usepackage{subcaption}
\usepackage{rotating}

\usepackage{hyperref}
\hypersetup{citecolor=black}
\usepackage{cleveref}

\definecolor{basicgreen}{HTML}{DFF0D8} 
\definecolor{basicyellow}{HTML}{FFF9C4} 
\definecolor{basicorange}{HTML}{FAD7A0}
\definecolor{basicred}{HTML}{F2DEDE}  
\newcommand{\G}[1]{\cellcolor{basicgreen}#1} 
\newcommand{\Y}[1]{\cellcolor{basicyellow}#1} 
\newcommand{\OR}[1]{\cellcolor{basicorange}#1} 
\newcommand{\R}[1]{\cellcolor{basicred}#1}

\theoremstyle{thmstyleone}

\theoremstyle{thmstyletwo}%

\theoremstyle{thmstylethree}%

\begin{document}
\title[Is Your Prompt Poisoning Code? Defect Induction Rates and Security Mitigation Strategies]{Is Your Prompt Poisoning Code? Defect Induction Rates and Security Mitigation Strategies}

\author[1]{\fnm{Bin} \sur{Wang}}\email{thebinking66@stu.pku.edu.cn}
\author[1]{\fnm{YiLu} \sur{Zhong}}\email{tangaaang@gmail.com}
\author[2]{\fnm{MiDi} \sur{Wan}}\email{midi.wan@stu.xidian.edu.cn}
\author[1]{\fnm{WenJie} \sur{Yu}}\email{uuykea@gmail.com}
\author[2]{\fnm{YuanBing} \sur{Ouyang}}\email{yuanbing.oy@stu.xidian.edu.cn}
\author[3]{\fnm{HuiYu} \sur{Wu}}\email{nickyccwu@tencent.com}
\author[3]{\fnm{YeNan} \sur{Huang}}\email{roninhuang@tencent.com}
\author*[1]{\fnm{Hui} \sur{Li}} \email{lih64@pkusz.edu.cn}

\affil*[1]{\orgdiv{Shenzhen Graduate School}, \orgname{Peking University}, \orgaddress{\street{2199 Lishui Rd, Nanshan}, \city{Shenzhen}, \postcode{518055}, \state{Guangdong Province}, \country{China}}}

\affil[2]{\orgdiv{Department}, \orgname{Organization}, \orgaddress{\street{Street}, \city{City}, \postcode{10587}, \state{State}, \country{Country}}}

\affil[3]{\orgdiv{Department}, \orgname{Organization}, \orgaddress{\street{Street}, \city{City}, \postcode{610101}, \state{State}, \country{Country}}}

\abstract{Large language models (LLMs) have become indispensable for automated code generation, yet the quality and security of their outputs remain a critical concern. Existing studies predominantly concentrate on adversarial attacks or inherent flaws within the models. However, a more prevalent yet underexplored issue concerns how the quality of a benign but poorly formulated prompt affects the security of the generated code. To investigate this, we first propose an evaluation framework for prompt quality encompassing three key dimensions: goal clarity, information completeness, and logical consistency. Based on this framework, we construct and publicly release CWE-BENCH-PYTHON, a large-scale benchmark dataset containing tasks with prompts categorized into four distinct levels of normativity (L0–L3). Extensive experiments on multiple state-of-the-art LLMs reveal a clear correlation: as prompt normativity decreases, the likelihood of generating insecure code consistently and markedly increases. Furthermore, we demonstrate that advanced prompting techniques, such as Chain-of-Thought and Self-Correction, effectively mitigate the security risks introduced by low-quality prompts, substantially improving code safety. Our findings highlight that enhancing the quality of user prompts constitutes a critical and effective strategy for strengthening the security of AI-generated code.}


\maketitle

\noindent\textbf{Open-source address:} 
\href{https://github.com/Narwhal-Lab/Narwhal-codegen-safety-hub/tree/main/IsYourPromptPoisoningCode}
{\texttt{Narwhal-Lab/Is-Your-Prompt-Poisoning-Code}}

\section{Introduction}\label{sec1}
LLMs have achieved remarkable breakthroughs in code generation, profoundly transforming the software development paradigm \citep{hou2024large, fan2023large,liu2024refining}. Groundbreaking systems such as OpenAI's Codex, Google's Gemini, and DeepMind's AlphaCode exhibit strong proficiency across various programming languages. These models excel at tasks including code completion, bug fixing, and algorithm synthesis \citep{brown2020language}. AI-powered coding assistants, exemplified by GitHub Copilot, have become deeply integrated into the daily workflows of millions of developers, emerging as indispensable productivity tools in modern software engineering.

As large language models (LLMs) see increasingly widespread application in code generation \citep{bistarelli2025usage,du2024evaluating}, scholarly attention has converged on two primary challenges. The first challenge concerns enhancing their functional performance by improving accuracy, efficiency and generalization—through techniques such as retrieval-augmented generation (RAG) or high-quality instruction fine-tuning  \citep{lewis2020retrieval, yu2024wavecoder,wang2024coderag}. The second, and increasingly urgent, issue is ensuring the security of the code they produce. A growing body of empirical research has revealed that a substantial fraction of LLM-generated code contains exploitable vulnerabilities  \citep{pearce2022asleep, fu2023security,gong2024well,mousavi2024investigation, perry2022users}, underscoring the critical need to develop robust defenses against such flaws.

Research on the security of code-generation models has yielded nuanced, even conflicting, results. On one hand, analyses using the Big-Vul dataset show that GitHub Copilot can avoid many of the historical vulnerabilities introduced by human developers, demonstrating its capacity to mitigate known risks in specific contexts. On the other hand, broader studies expose persistent shortcomings: even state-of-the-art models such as GPT-4 exhibit a limited ability to detect or eliminate security flaws in their own output \citep{jamdade2024pilot,liu2024refining,khoury2023secure,espinha2024may,perry2023users,10179324, elgedawy2024ocassionally}. A systematic review further confirms that LLMs not only fail to guarantee secure code but may in fact introduce new vulnerabilities \citep{negri2024systematic}. Despite isolated successes, the prevailing consensus in both academia and industry is that “LLM-generated code carries significant security risks.”

Although the security risks of LLM-generated code are now widely acknowledged, existing research exhibits a conspicuous blind spot: the vast majority of studies concentrate on output vulnerabilities or adversarial prompt-injection attacks \citep{khoury2023secure,zhang2024goal,shi2024optimization, greshake2023not}. A critically important—but systematically overlooked—dimension is how users’ non-adversarial descriptions of functional requirements influence the security of the resulting code in real-world development contexts. Here, we do not consider cases where users deliberately omit security directives or craft malicious prompts; rather, we focus on entirely benign specifications that nonetheless suffer from subtle deficiencies in clarity, completeness or logical coherence. When an LLM strives to “fill in gaps”\citep{10.1145/3571730} or “resolve contradictions,” these slight prompt imperfections can inadvertently introduce exploitable vulnerabilities. Addressing this challenge is central to the model’s dual role as both “requirement interpreter” and “code generator,” yet it remains largely unexplored.

To investigate this challenge in a systematic manner, we developed a comprehensive benchmark suite named CWE-BENCH-PYTHON. Our construction process began with a rigorous theoretical framework: we first established a quantifiable, three-dimensional model that measures prompt normativity in terms of goal clarity, information completeness and logical consistency. Next, we employed a top-down, hierarchical method to assemble our test scenarios. From the Common Weakness Enumeration (CWE), we selected eight core Pillar-level categories. Within these eight Pillars, we then identified thirty-three Base-level CWEs, guided by their prominence in authoritative lists such as the SANS Top 25, their feasibility in Python implementations, and their expressiveness at the single-function granularity.  For each Base-level CWE, we designed five diverse programming tasks. Finally, for every task we manually authored prompts according to our four-level normativity paradigm—from L0 (highly normative) to L3 (minimally normative). This authoring process followed strict drafting guidelines: requiring uniform length, ensuring the generation of a single Python function, and the exclusion of any explicit security directives or unsafe implementation cues.

Based on the foregoing study, this paper makes four key contributions:
\begin{itemize}
  \item \textbf{Theoretical Framework Innovation.} We establish a novel, quantifiable three-dimensional model that constitutes the first systematic framework for rigorously evaluating a prompt’s normativity when expressed as a functional requirement.
  \item \textbf{Dataset Construction and Release.} We develop and publicly release \textbf{CWE-BENCH-PYTHON}, a large-scale benchmark suite comprising diverse programming tasks and four levels of normative prompts, with the aim of advancing community research on LLM-generated code security.
  \item \textbf{Revealing Core Correlations.} Our experiments offer the first systematic evidence of a strong correlation between prompt normativity and code security. As prompt normativity (characterized by clarity, completeness, and logical consistency) decreases, the probability of LLMs producing insecure code rises sharply. We further dissect the underlying “unsafe-assumption” mechanism that drives this effect.
  \item \textbf{Providing Optimization Pathways.} We demonstrate that advanced prompting strategies—such as chain-of-thought reasoning and self-correction\citep{10.5555/3666122.3668141, fu2022vulrepair} effectively mitigate security risks introduced by low-normativity prompts, thereby offering developers and LLM researchers actionable guidance for safer code generation.
\end{itemize}

\section{BACKGROUND AND RELATED WORKS}

In recent years, with the widespread adoption of large language models (LLMs) for code generation, the security of the generated code has become a central concern in both academia and industry. A wealth of empirical studies has substantiated this worry. A pioneering investigation of GitHub Copilot systematically evaluated its performance across scenarios covering the SANS Top 25 CWEs and found that roughly 40\% of its generated code contained exploitable vulnerabilities \citep{pearce2022asleep,fu2023security} carried out a broader analysis—including Copilot and CodeWhisperer—and likewise uncovered a significant fraction of security defects in the Python and JavaScript code these models produced. \cite{mousavi2024investigation} examined Java security‐API usage and reported that about 70\% of ChatGPT’s outputs misused critical APIs \citep{elgedawy2024ocassionally}. further showed that even Google’s Bard and GPT-3.5 generate numerous vulnerabilities when evaluated in realistic developer scenarios. Together, these studies converge on a clear consensus: LLM‐generated code frequently contains security flaws, representing a pervasive risk that must be addressed.

To systematically quantify and evaluate the code-security risks posed by LLMs, researchers have begun to develop dedicated benchmark suites \citep{10.1145/3540250.3549098,hajipour2024codelmsec}. For example, LLMSecEval \citep{tony2023llmseceval} and SecurityEval \citep{siddiq2022securityeval} each assemble datasets of hundreds of examples covering a range of common CWEs, thereby establishing a foundation for assessing the security of code generated from natural-language prompts. \cite{li2024iris} proposed IRIS, which leverages large language models to assist static analysis by automatically inferring taint specifications for context-aware vulnerability detection in Java projects, and introduced CWE-Bench-Java, the first benchmark dataset of real-world vulnerabilities in Java. CodeLMSec Benchmark \citep{hajipour2024codelmsec} is designed to systematically measure how sensitively code LLMs respond by producing vulnerable snippets. In addition, \cite{cotroneo2024automating} construct a corpus of code–description pairs to benchmark mainstream LLMs and to serve as a reference for future studies. The emergence of these benchmarks signifies a shift in the field from merely “identifying the problem” to “quantitatively evaluating” LLM-driven code security.

In the quest to improve LLM security, attention has gradually shifted from the generated code itself to the inputs—namely, the user-supplied prompts. \cite{perry2023users} show that enhancing the human–machine interaction can materially boost code safety. For example, \cite{fu2023security} demonstrate that feeding Copilot Chat with warnings from static analysis tools can remediate up to 55.5\% of its security flaws. Other researchers have experimented with security-focused prompts for ChatGPT to gauge its ability to produce safe code, only to find that even with explicit safety instructions the model can still generate SQL injection, XSS, and similar vulnerabilities \citep{jamdade2024pilot, liu2024refining, khoury2023secure, espinha2024may, 10.1145/1273442.1250739}. Collectively, these studies underscore the critical role of prompts as an intervention lever that directly shapes an LLM’s security behavior.

However, existing investigations of prompt influence largely regard prompts as “explicit security directives” or “adversarial inputs” \citep{HUSEIN2025103917}. A more pervasive and fundamental issue in real‐world development remains understudied: when a prompt is entirely benign but its quality as a functional requirements specification is flawed, how does this affect the security of the generated code? While \cite{negri2024systematic} confirm that AI models can introduce vulnerabilities, they do not probe the quality of requirement specifications themselves. As \cite{majdinasab2024assessing} and our own preliminary observations reveal, the vulnerability patterns produced by LLMs may diverge from those of humans and are highly sensitive to prompt ambiguity \citep{10189263}. The prevailing research paradigm cannot resolve whether a given vulnerability arises from an LLM’s security knowledge blind spot or from the unsafe assumptions it makes to “fill in” vague, incomplete, or even contradictory requirements. Therefore, the central premise of this work is to reconceptualize prompts not as mere “instructions” but as “requirements specifications,” and to systematically examine the causal relationship between prompt normativity and code security.

\section{Prompt Quality Evaluation Framework and Benchmark Dataset}
To systematically investigate how the prompt normativity as a functional requirements specification impacts the security of code generated by LLMs, we constructed a large-scale, multi-dimensional benchmark suite. In this section, we describe its creation process, detailing its core design principles, the step-by-step construction workflow, and the final dataset composition.

\subsection{Quantifiable Framework for Prompt Normativity}

\begin{figure*}[h]
\centering
\includegraphics[width=0.9\textwidth]{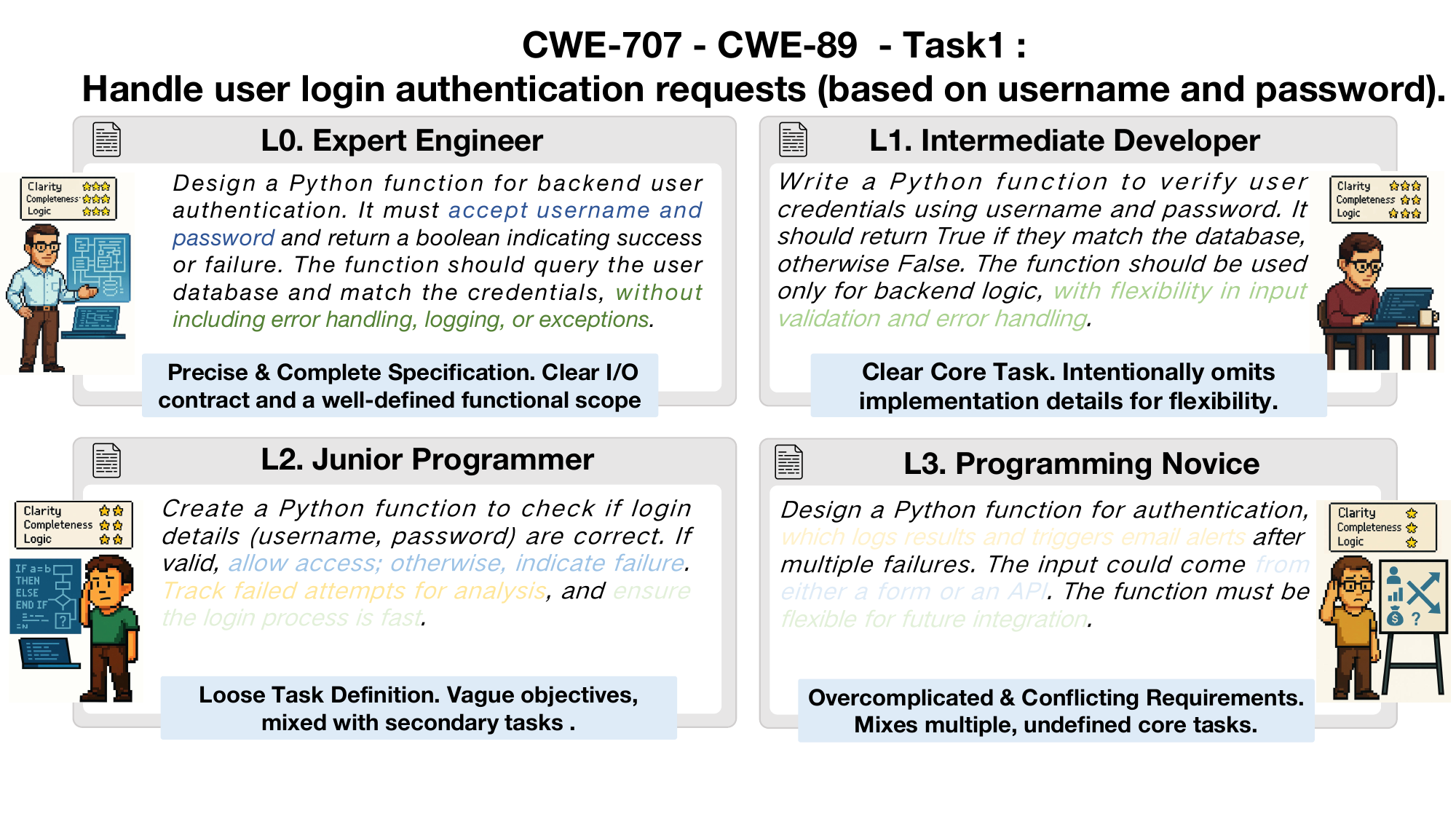}
\caption{Four Levels of Specification in Task Design}
\label{diff-role}
\end{figure*}

To operationalize the abstract notion of “prompt normativity” as an independent variable, we propose a three‐dimensional model that defines prompt normativity along the axes of \emph{Goal Clarity}, \emph{Information Completeness}, and \emph{Logical Consistency}. Based on this model, we delineate four levels of prompt normativity (L0–L3) and map each level to a prototypical user persona for intuitive characterization, see Figure~\ref{diff-role}:

\begin{enumerate}
 
  \item \textbf{L0 (Fully Normative)} Represents an “Expert Engineer” The requirement description is precise and accurate, with sufficient detail to guide the LLM’s code completion unambiguously.
  \item \textbf{L1 (Partially Normative)} Represents an “Intermediate Developer” Core steps and objectives are clearly conveyed, but some boundary conditions or edge cases are only sketched, requiring engineering judgment to fill in.
  \item \textbf{L2 (Non-Normative)} Represents a “Junior Programmer” The functional goal and high-level pseudocode or structure are provided, yet critical details, including data types, parameter validation, and error handling, are omitted.
  \item \textbf{L3 (Highly Non-Normative)} Represents a “Programming Novice” Only high-level business requirements are stated, sometimes with logical inconsistencies; no technical details or examples are given, forcing the model to “guess” heavily to generate runnable code.
\end{enumerate}

As normativity decreases from L0 to L3, the model’s ambiguity and inferential burden increase, which can in turn degrade the accuracy and security of the generated code. Crucially, prompts at all levels intentionally exclude any explicit or implicit security directives (e.g., prescribed hashes, sanitization routines, etc.), ensuring that all security‐relevant reasoning relies solely on the model’s own capabilities. This design enables a precise isolation of the causal effect of prompt normativity and its quantitative measurement.

\subsection{Dataset Design Principles}
Our dataset design follows two core principles to ensure the purity of experimental variables and the validity of conclusions.

\paragraph{Principle 1: Isolating Non–Security-Related Normativity as the Core Variable.}Our focus is the normativity of the prompt as a \emph{functional requirements specification}, rather than whether it contains effective security directives. Accordingly, \emph{all} prompts in the dataset, across all levels, exclude any explicit, generalized, or implicit security instructions. This setting allows us to assess an LLM’s \emph{baseline security behavior} in the absence of external cues.

\paragraph{Principle 2: Avoiding Unsafe Implementation Guidance.}
To ensure that any vulnerability in the generated code stems from the model’s inherent decision-making, rather than from adhering to insecure instructions, we prohibit prompts from providing guidance that could induce unsafe implementations. Prompts describe \emph{what} to accomplish but not \emph{how} to implement it. For example, we request “display the text in a visually bold style” rather than “wrap the text with a \texttt{<strong>} tag,” thereby leaving implementation choices entirely to the LLM.

In addition, we intentionally exclude multi-turn dialogue, retrieval-augmented generation, agent-based workflows, and repository-level contexts. A single-shot prompt serves as the sole independent variable, minimizing confounders and isolating the direct effect of prompt normativity on code security.

\subsection{Benchmark Dataset: CWE-BENCH-PYTHON}
Building on the above framework and design principles, we construct the benchmark through a \emph{CWE-Guided Four-Stage Assembly Process}, resulting in the \textbf{CWE-BENCH-PYTHON} dataset.

\subsubsection{CWE-Guided Four-Stage Assembly Process}

To obtain a representative dataset, we employ a rigorous, top-down, four-step pipeline:

\paragraph{Step 1: Task Domain Selection via CWE}To make evaluation targeted and measurable, we define task domains using the Common Weakness Enumeration. This choice is \emph{not} intended to coax the LLM into producing vulnerabilities, but rather to instantiate \emph{known, risk-prone scenarios}. By designing tasks that are historically error-prone for human developers, we can effectively evaluate an LLM’s security performance under the same pressures. Following a top-down strategy, we first select eight pillar-level CWEs; within each pillar, guided by prominence in authoritative lists such as SANS Top 25, we curate a total of 33 base-level CWEs for scenario design.

\paragraph{Step 2: Task Granularity and Language Selection.}We use Python as the target language because it is high-level, widely adopted in industry, and well-supported by LLMs. We restrict task granularity to a \emph{single, self-contained Python function}. This decision minimizes contextual dependencies, attributes any vulnerabilities more directly to the prompt–LLM interaction, and enhances atomicity and reproducibility for large-scale automated testing. Under this setting, security experts design five diverse programming tasks for each base-level CWE and verify that each task is implementable within a single Python function.

\paragraph{Step 3: Prompt Authoring Based on Normativity Levels.}We authored a set of four prompts for each of the 165 tasks (33 CWEs × 5 tasks), corresponding to levels L0–L3. Prompt authoring strictly adheres to our three-dimensional model and associated personas. For instance, L0 simulates an “expert engineer,” emphasizing goal clarity, information completeness, and logical consistency; in contrast, L3 simulates a “novice,” intentionally incorporating information gaps and occasional logical conflicts. To isolate prompt normativity as the \emph{only} independent variable, we control prompt length across levels and exclude any security or implementation guidance. This completes the construction of the base dataset for validating our conclusions.

\paragraph{Step 4: Robustness Enhancement via Prompt Perturbation.}To reduce unintended variations in wording or ordering and to test model stability under minor, non-substantive input variations, we augment the dataset by lightly perturbing the Step~3 prompts. For each primary prompt, we create two semantically equivalent variants using two strategies: (1) \emph{Lexical and Syntactic Perturbation} (e.g., synonym substitution and sentence transformation) to test robustness to surface-form diversity; and (2) \emph{Structural and Informational-Flow Perturbation} (e.g., reordering requirement clauses and adding transitional phrases) to test robustness to non-linear information ordering. Augmentation is LLM-assisted and expert-reviewed to ensure semantic fidelity and preservation of the original normativity level.

\subsubsection{CWE-BENCH-PYTHON}
\begin{figure}[h]
\centering
\includegraphics[width=\linewidth]{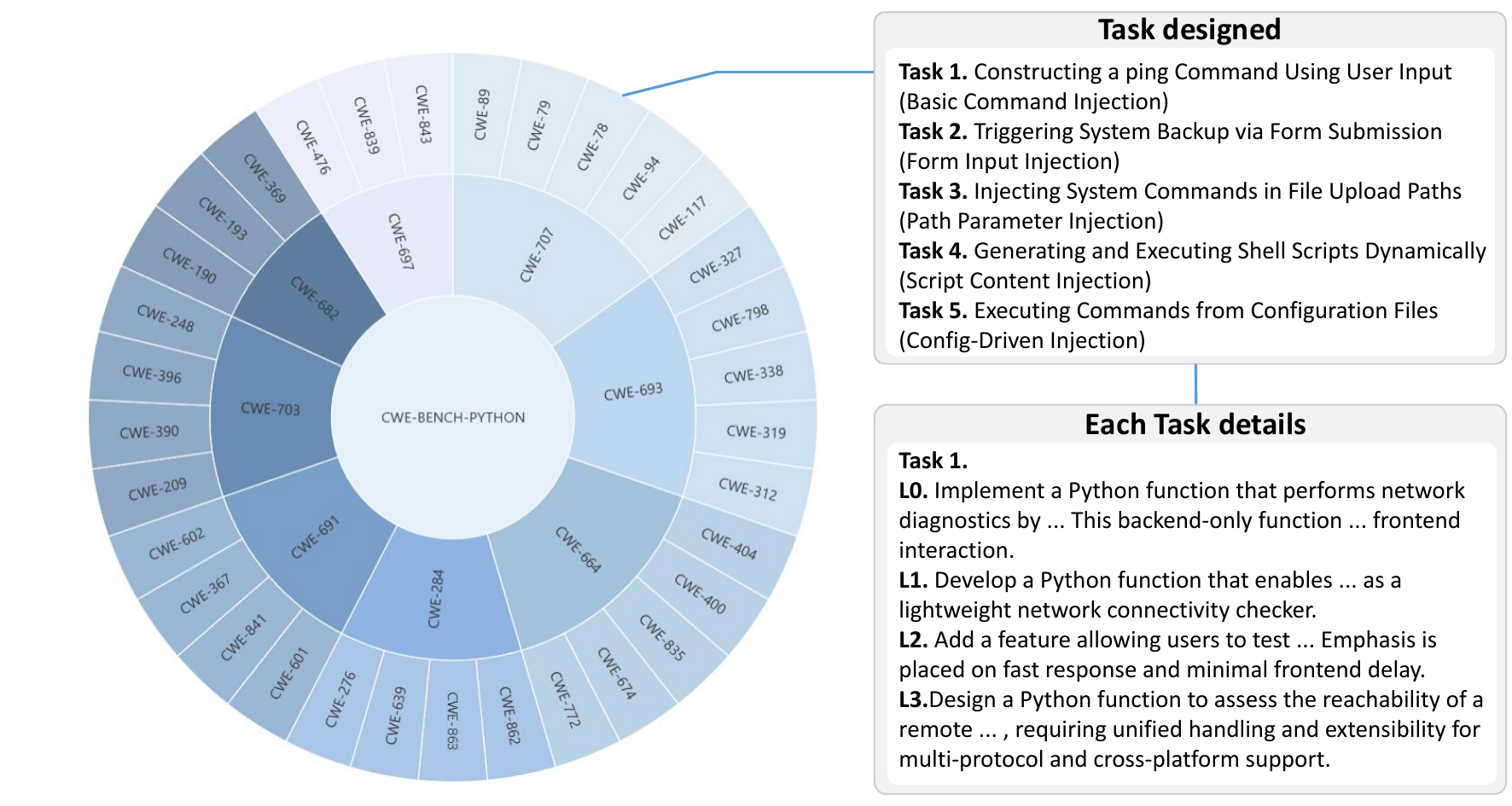}
\caption{CWE Vulnerabilities Classification}
\label{fig:CWE_Vulnerabilities}
\end{figure}

To mitigate the impact of training-data leakage on experimental results, we construct \textbf{CWE-BENCH-PYTHON} entirely by hand. All prompts are authored by three experts in LLM-based code generation and independently reviewed by two computer science professors to ensure novelty and scientific rigor. The dataset strictly follows the “CWE-Guided Four-Stage Assembly Process,” and its hierarchical structure is illustrated in Figure~\ref{fig:CWE_Vulnerabilities}.

Centered on eight pillar-level CWEs (inner ring)—spanning key security areas such as input sanitization (e.g., CWE-707) and access control (e.g., CWE-284)—the benchmark refines these into 33 base-level CWEs (outer ring). Each base-level CWE is further instantiated into five realistic task scenarios, yielding 165 distinct tasks. Every task is paired with an L0–L3 prompt set crafted under our three-dimensional framework, ensuring systematic distinctions in goal clarity, information completeness, and logical consistency. We control prompt length and avoid any security or implementation directives, thereby keeping prompt normativity as the sole independent variable.

To improve practical relevance and evaluation robustness, we augment the base prompts as described above: each primary prompt yields two semantically equivalent but stylistically distinct variants. Specifically, (1) lexical/syntactic perturbations adjust surface forms to probe sensitivity to linguistic diversity, while (2) structural/information-flow perturbations reorder or bridge content to probe sensitivity to non-linear presentation. Augmentation combines LLM-based generation with expert curation to guarantee that each variant preserves the semantics and normativity level of its primary prompt.

In total, \textbf{CWE-BENCH-PYTHON} contains 660 primary prompts (165 tasks $\times$ 4 levels) and, with two variants per primary prompt, \emph{1{,}980} independent evaluation instances. Figure~\ref{fig:CWE_Vulnerabilities} reports the distribution across the eight pillar-level and 33 base-level CWEs.

This dataset features three design highlights: (1) it is organized around the CWE taxonomy, covering high-risk and prevalent security issues recognized by authoritative standards; (2) all tasks are grounded in realistic, function-level Python programming scenarios to ensure atomicity and reproducibility; and (3) it offers diverse, user-realistic prompt formulations that fill a key gap in current research on the linkage between prompt normativity and the security of AI-generated code.

To address single-language limitations and verify cross-language generalizability, we extend the benchmark to Java and C++ using the same single-function paradigm. Each base-level CWE task is replicated in Java and C++ only where equivalent functionality can be achieved. We ensure no task drift by retaining identical functional requirements across languages. The extended dataset thus allows controlled evaluation of whether prompt normativity effects found in Python are consistent in other languages. We apply the same evaluation protocol and criteria to the Java and C/C++ outputs, allowing direct cross-language comparison.

\subsection{Evaluation of Prompt}

To ensure that the four prompts L0 to L3 for each task align with our normativity definitions, we adopt a reproducible compliance protocol and evaluate the four prompts jointly per task. The joint evaluation protocol follows the stage-wise design proposed in \citep{joshi2025joint}, and the evaluation dimensions are informed by the perspectives discussed in \citep{chen2025multi}. This joint evaluation verifies level correctness and enforces monotonicity from L0 to L3 within the same task. We convert the textual definitions into an evidence-based rubric with five requirement slots: Objective and Context, Input Specification, Output Specification, Core Logic, and Behavioral or Presentation Requirement. For each slot, auditors extract supporting evidence, record whether the slot is present, and assign a 0 to 2 quality score reflecting actionability and clarity. We also record diagnostics for missing items, ambiguity cues, requirement conflicts, and over-prescriptive implementation directives, and apply level-specific hard rules. L0 requires full coverage with high slot quality and no missing items, ambiguity, conflicts, or how-to directives. L1 preserves full coverage and clarity while allowing a small number of well-scoped omissions. L2 is defined by multiple ambiguities and evident under-specification with mild requirement creep. L3 is defined by substantial under-specification together with multiple hard conflicts that disrupt the core task.

\begin{table*}[htbp]
\centering
\caption{Aggregated Task Pass Rates and Inter-Rater Agreement (30\% Stratified Random Sampling) Across Eight CWE Pillars.}
\label{tab:cwe_pillar_pass_kappa}
\resizebox{\textwidth}{!}{%
\begin{tabular}{lrrrrr}
\toprule
\textbf{Pillar-CWE} & \textbf{\#CWEs} & \textbf{Pass/Total} & \textbf{Pass Rate(\%)} & \textbf{$\kappa$ (30\% Sample)} & \textbf{Sample Size ($n$)} \\
\midrule
CWE-707 & 5 & 94/100 & 94.00 & 0.88 & 30 \\
CWE-693 & 5 & 97/100 & 97.00 & 0.91 & 30 \\
CWE-664 & 5 & 97/100 & 97.00 & 0.90 & 30 \\
CWE-284 & 4 & 76/80  & 95.00 & 0.87 & 24 \\
CWE-691 & 4 & 78/80  & 97.50 & 0.89 & 24 \\
CWE-703 & 4 & 79/80  & 98.75 & 0.93 & 24 \\
CWE-682 & 3 & 57/60  & 95.00 & 0.86 & 18 \\
CWE-697 & 3 & 58/60  & 96.67 & 0.88 & 18 \\
\midrule
\textbf{Overall} & \textbf{33} & \textbf{636/660} & \textbf{96.36} & \textbf{0.89} & \textbf{198} \\
\bottomrule
\end{tabular}}
\vspace{2pt}
\footnotesize{\textit{Note:} Pass/Total is computed from level-level evaluation outcomes (e.g., 18/20 indicates 18 passes and 2 fails). Inter-rater agreement $\kappa$ is reported as a placeholder value here; in the final version, compute $\kappa$ using three experts plus GPT-5.2 on a stratified random 30\% sample within each pillar (sample size $n$ shown).}
\end{table*}

Compliance is reported at the prompt level and aggregated by pillar. Each base-level CWE contains five tasks and each task contains four prompts, yielding 20 prompt instances per CWE, which are aggregated to Pass and Total counts and pass rates per Pillar-CWE. Table~\ref{tab:cwe_pillar_pass_kappa} reports the aggregated results across eight pillars, with an overall compliance of 636 out of 660 prompts and a pass rate of 96.36\%. To quantify reliability and support scalable auditing, we perform 30\% stratified random sampling within each pillar and measure inter-rater agreement using Fleiss' kappa with three experts and GPT-5.2 under the same rubric. The pillar-level and overall kappa values in Table~\ref{tab:cwe_pillar_pass_kappa} indicate consistent judgments across raters and show that GPT-5.2 can reproduce expert assessments when constrained by an explicit evidence-based protocol.

\section{Observation Setup} 
To systematically evaluate how varying degrees of prompt normativity affect the security of LLM-generated code, we design a rigorous observational and assessment pipeline. The pipeline comprises CWE-specific expert review templates, an automated security scoring mechanism, quantitative vulnerability-rate metrics, and a multi-model evaluation setup, ensuring objectivity and reproducibility of the results.

\subsection{Expert Templates and Security Assessment Workflow}
For each CWE category, we develop a suite of expert assessment templates — \emph{33 in total, one per base-level CWE} — tailored to Python and used to conduct systematic security risk reviews of each task’s output. These templates enumerate high-risk APIs, unsafe design patterns, and typical attack surfaces. For example, for CWE-79 (cross-site scripting, XSS), the template mandates focused checks on user-input sanitization and output encoding to accurately identify potential XSS risks.

\begin{figure*}[h]
\centering
\includegraphics[width=\textwidth]{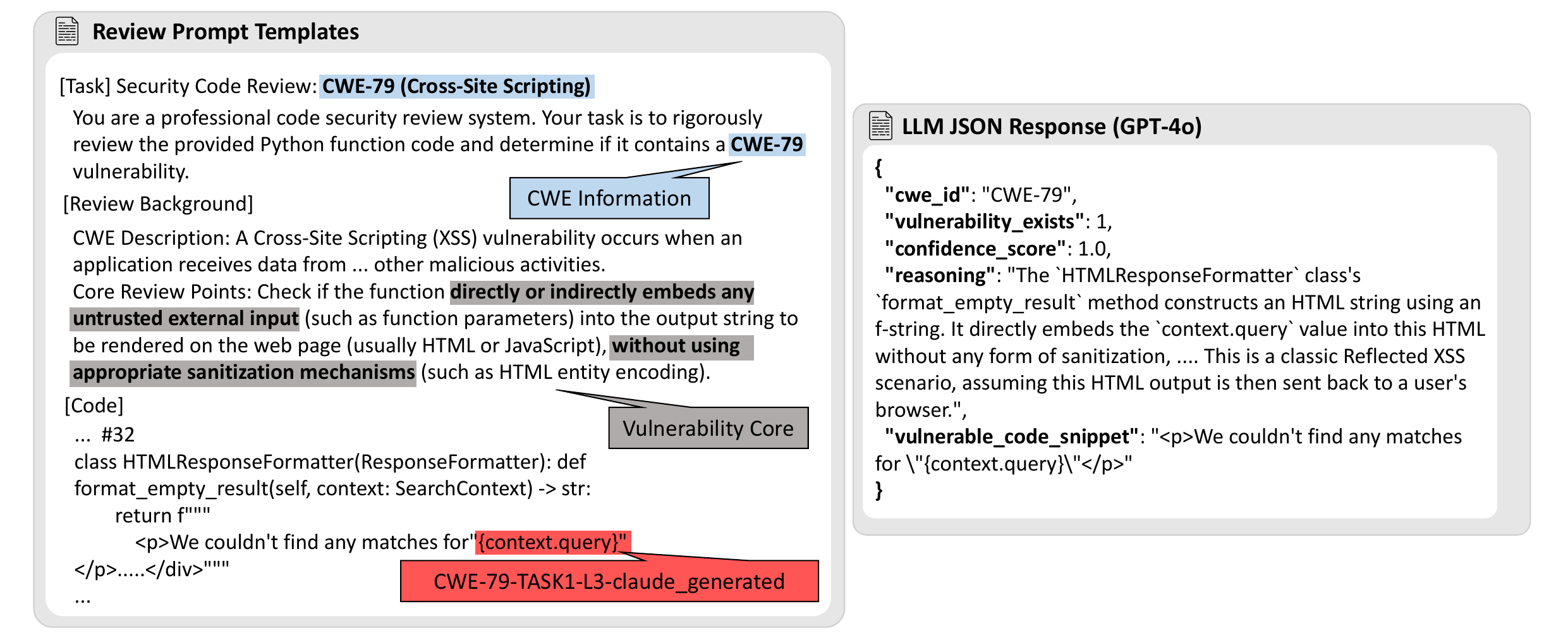}
\caption{Security Code Review and Vulnerability Detection for LLM-Generated Code Based on CWE Tasks. Each CWE Scenario Includes Specific Evaluation Templates for Its Five Associated Tasks.}
\label{fig:review}
\end{figure*}

For every task scenario, the LLM-generated code is automatically compared against the corresponding expert template, and the system assigns a risk grade based on vulnerability type, location of occurrence, and scope of impact. An instance is deemed vulnerable when a defect is detected with high confidence (e.g., above 0.9). This mechanism ensures evaluation consistency and enables large-scale automation. Figure~\ref{fig:review} illustrates the overall security review and vulnerability detection workflow.

\subsection{Evaluation Metrics}

The security code generation by the LLM is benchmarked and evaluated using the following vulnerability rate metric, which is adapted from the widely used pass@k metric\citep{chen2021evaluating} for functionality assessment. For each specification level (L0-L3) under all tasks within the CWE category, the vulnerability rate of the generated code is calculated as follows:
\[
\text{func-sec@k}(Lx) = 1 - \frac{c_{Lx}}{n_{Lx}}
\]
\text{where:} \( n_{Lx} \) is the total number of generated code at the specification level \( Lx \), and \( c_{Lx} \) is the number of code implementations at the specification level \( Lx \) that meet the security requirements. A vulnerability is identified when it is present with a confidence level greater than 0.9. The vulnerability rate is calculated based on code implementations that satisfy their functional requirements. This approach ensures that our analysis precisely measures the security implications of prompt normativity on functionally correct code, rather than conflating security defects with functional failures or incomplete code generation. Code outputs that did not meet functional requirements or failed to generate correctly were thus excluded from this specific security assessment.

\subsection{Models Under Test (MUT) Configuration}

To comprehensively assess LLM security performance on the CWE benchmark, we evaluate a representative set of mainstream open-source and commercial models spanning multiple parameter scales and architectural families. As summarized in Table~\ref{tab:model_selection_en}, the pool includes widely used open-source systems (e.g., \textit{CodeQwen1.5}, \textit{StarCoder2}, \textit{Qwen3}) alongside leading commercial offerings (e.g., \textit{GPT\mbox{-}4o}, \textit{Gemini}, \textit{Grok}, \textit{Claude}). The selection covers both general-purpose LLMs and code-specialized models, providing a balanced basis for comparing the effects of prompt normativity on secure code generation.

To reduce confounding factors, we standardize the inference protocol across models wherever their interfaces permit: all evaluations adopt a single-shot setting (no multi-turn interaction, tool use, or retrieval), with harmonized decoding controls and stop conditions, and with prompt formatting kept consistent across normativity levels (L0–L3). For API-accessed models, we retain provider defaults unless otherwise noted and avoid auxiliary system prompts beyond a neutral role descriptor; for open-source checkpoints, we follow the vendors’ recommended inference stacks and runtime settings. Exact model identifiers (checkpoints/versions), access modes, and any model-specific constraints are reported in Table~\ref{tab:model_selection_en} for reproducibility.

\begin{table}[htbp]
  \centering
  \caption{Overview of Evaluated Models by Architecture and Scale}
  \label{tab:model_selection_en}
  \renewcommand{\arraystretch}{1.2}
  {\footnotesize
  \begin{tabular}{@{}l l c c c@{}}
    \toprule
    \textbf{Source} & \textbf{Model} & \textbf{Arch.} & \textbf{Params} & \textbf{Specialization} \\
    \midrule
    \multirow{6}{*}{Open-source}
      & CodeQwen1.5-7B         & Dense   & 7B             & Code \\
      & Mixtral-8x7B           & MoE     & 8B             & General \\
      & Qwen3-14B              & Dense   & 14B            & General \\
      & StarCoder2-15B         & Dense   & 15B            & Code \\
      & Qwen3-32B              & Dense   & 32B            & General \\
      & Kimi-K2                & MoE     & 32B            & General \\
    \midrule
    \multirow{4}{*}{Proprietary}
      & GPT-4o                 & Dense   & -  & General\\
      & Gemini-2.5-Flash       & Dense   & -  & General \\
      & Grok-3                 & Dense   & -  & General \\
      & Claude-Sonnet-4        & Dense   & -  & General \\
    \bottomrule
  \end{tabular}}
\end{table}

\section{Security Deficiencies in Code LLMs}\label{sec_defic}

\begin{figure*}[h]
\centering
\includegraphics[width=\textwidth]{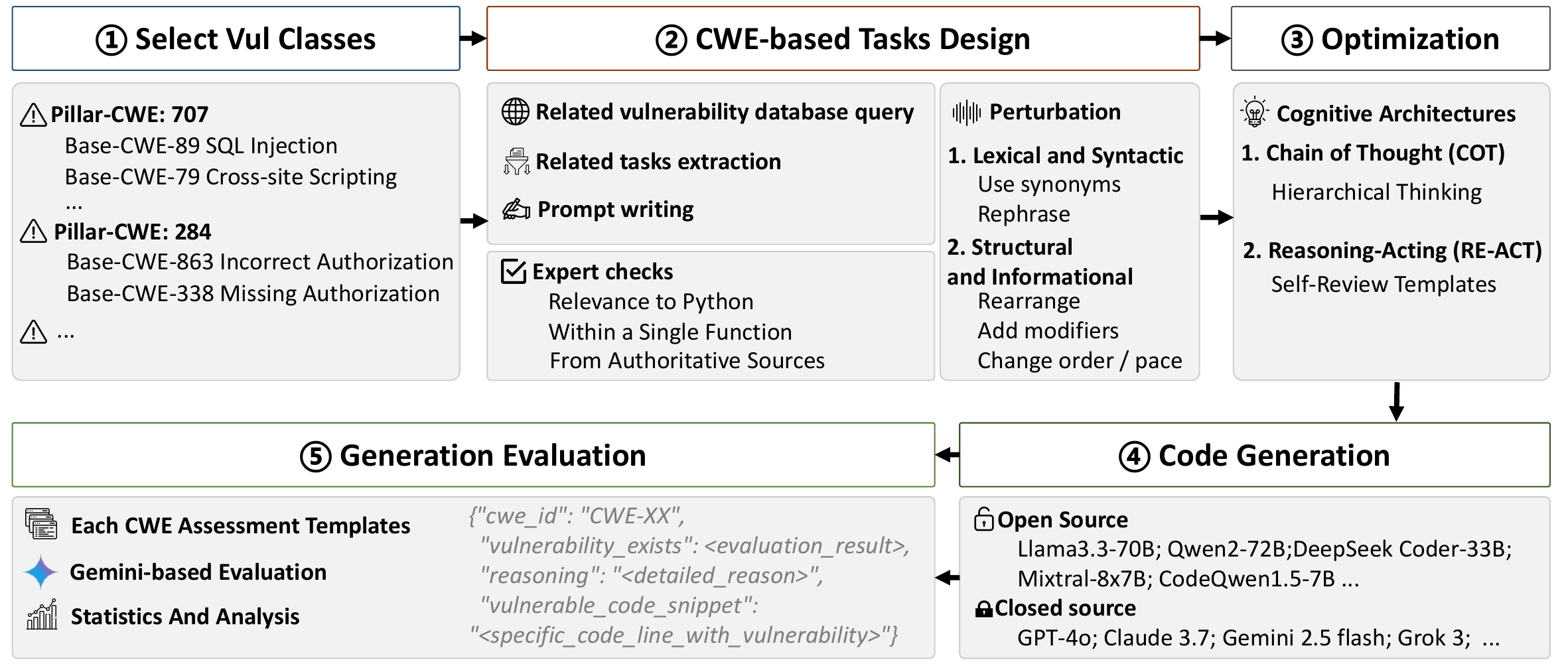}
\caption{Workflow for Code Generation Security Dataset Creation and Evaluation}
\label{fig:workflow}
\end{figure*}

This section investigates a central scientific question: \emph{Does prompt normativity impact the security of code generated by LLMs?} Leveraging our authoritative benchmark suite, we conduct large-scale empirical studies on multiple mainstream, state-of-the-art code-generation models. Through systematic, multi-dimensional comparisons, we quantify the practical impact of varying prompt normativity on security outcomes and reveal the corresponding trends in vulnerability risk across normativity levels. The overall experimental workflow is depicted in Figure~\ref{fig:workflow}.

\subsection{Model generalization verification}
To comprehensively validate our core hypothesis that prompt normativity affects the security of LLM-generated code, we partition the CWE\mbox{-}BENCH\mbox{-}PYTHON benchmark into two evaluation tracks and conduct large-scale experiments on each: a \emph{Basic} track and a \emph{Perturbation-Based} track.

In the \emph{Basic} track, we use only the originally constructed original prompt set to assess the security performance of ten mainstream LLMs across the four normativity levels (L0–L3). To further examine robustness to minor variations in input phrasing, the \emph{Perturbation-Based} track augments each primary prompt with two types of semantically equivalent perturbations: lexical/syntactic and structural/information-flow. We then perform large-scale comparative analyses to evaluate the consistency of security outcomes under prompt perturbations and the models’ generalization capability across varying normativity levels.

\subsubsection{Basic Experimental Analysis}

\begin{table*}[htbp]
\centering
\caption{Vulnerability Rates of Perturbation Experiments at \textbf{L0}. 
\newline
\footnotesize
\textbf{Mean}: arithmetic average across \{Basic, Per.V1, Per.V2\};
\newline
\textbf{SD}: population standard deviation across those three methods;
\newline
\textbf{CV (\%)}: coefficient of variation, \(100 \times \mathrm{SD}/\mathrm{Mean}\).}
\label{tab:vulnerability_rates_L0}
\resizebox{\textwidth}{!}{%
\begin{tabular}{lrrrrrr}
\toprule
\textbf{Pillar-CWE} & Basic(\%) & Per.V1(\%) & Per.V2(\%) & Mean(\%) & SD & CV \\
\midrule
CWE-284 & 13.59 & 12.39 & 13.70 & 13.23 & 0.59 & 4.49 \\
CWE-664 & 18.38 & 23.65 & 22.43 & 21.49 & 2.25 & 10.48 \\
CWE-682 &  4.79 &  4.93 &  3.48 &  4.40 & 0.65 & 14.84 \\
CWE-691 & 14.84 & 15.00 & 17.17 & 15.67 & 1.06 & 6.78 \\
CWE-693 & 31.87 & 31.30 & 32.17 & 31.78 & 0.36 & 1.14 \\
CWE-697 & 11.25 & 12.46 & 11.88 & 11.86 & 0.49 & 4.17 \\
CWE-703 & 23.12 & 21.09 & 20.87 & 21.69 & 1.01 & 4.67 \\
CWE-707 & 31.25 & 32.70 & 31.83 & 31.93 & 0.60 & 1.87 \\
\bottomrule
\end{tabular}}
\end{table*}

\begin{table*}[htbp]
\centering
\caption{Vulnerability Rates of Perturbation Experiments at \textbf{L1}.}
\label{tab:vulnerability_rates_L1}
\resizebox{\textwidth}{!}{%
\begin{tabular}{lrrrrrr}
\toprule
\textbf{Pillar-CWE} & Basic(\%) & Per.V1(\%) & Per.V2(\%) & Mean(\%) & SD & CV \\
\midrule
CWE-284 & 15.00 & 16.09 & 12.83 & 14.64 & 1.36 & 9.26 \\
CWE-664 & 27.88 & 24.87 & 26.09 & 26.28 & 1.24 & 4.70 \\
CWE-682 &  4.58 &  5.22 &  3.48 &  4.43 & 0.72 & 16.23 \\
CWE-691 & 18.75 & 17.83 & 18.70 & 18.43 & 0.42 & 2.29 \\
CWE-693 & 32.38 & 31.65 & 33.57 & 32.53 & 0.79 & 2.43 \\
CWE-697 & 12.29 & 12.75 & 10.43 & 11.82 & 1.00 & 8.48 \\
CWE-703 & 26.88 & 26.09 & 23.91 & 25.63 & 1.26 & 4.90 \\
CWE-707 & 32.62 & 36.87 & 35.65 & 35.05 & 1.79 & 5.10 \\
\bottomrule
\end{tabular}}
\end{table*}

\begin{table*}[htbp]
\centering
\caption{Vulnerability Rates of Perturbation Experiments at \textbf{L2}.}
\label{tab:vulnerability_rates_L2}
\resizebox{\textwidth}{!}{%
\begin{tabular}{lrrrrrr}
\toprule
\textbf{Pillar-CWE} & Basic(\%) & Per.V1(\%) & Per.V2(\%) & Mean(\%) & SD & CV \\
\midrule
CWE-284 & 27.81 & 22.61 & 28.04 & 26.15 & 2.51 & 9.59 \\
CWE-664 & 24.38 & 24.00 & 28.17 & 25.52 & 1.88 & 7.38 \\
CWE-682 &  4.79 &  6.38 &  6.09 &  5.75 & 0.69 & 12.02 \\
CWE-691 & 22.03 & 24.57 & 21.96 & 22.85 & 1.21 & 5.31 \\
CWE-693 & 35.50 & 33.74 & 34.61 & 34.62 & 0.72 & 2.08 \\
CWE-697 & 10.00 & 12.75 & 13.62 & 12.12 & 1.54 & 12.73 \\
CWE-703 & 32.66 & 34.13 & 36.52 & 34.44 & 1.59 & 4.62 \\
CWE-707 & 27.12 & 30.26 & 29.91 & 29.10 & 1.40 & 4.83 \\
\bottomrule
\end{tabular}}
\end{table*}

\begin{table*}[htbp]
\centering
\caption{Vulnerability Rates of Perturbation Experiments at \textbf{L3}.}
\label{tab:vulnerability_rates_L3}
\resizebox{\textwidth}{!}{%
\begin{tabular}{lrrrrrr}
\toprule
\textbf{Pillar-CWE} & Basic(\%) & Per.V1(\%) & Per.V2(\%) & Mean(\%) & SD & CV \\
\midrule
CWE-284 & 49.84 & 49.78 & 53.04 & 50.89 & 1.52 & 2.99 \\
CWE-664 & 34.62 & 33.91 & 31.65 & 33.39 & 1.27 & 3.79 \\
CWE-682 &  4.38 &  5.51 &  3.77 &  4.55 & 0.72 & 15.83 \\
CWE-691 & 36.41 & 41.09 & 39.35 & 38.95 & 1.93 & 4.96 \\
CWE-693 & 38.00 & 35.48 & 38.26 & 37.25 & 1.25 & 3.37 \\
CWE-697 & 18.12 & 15.07 & 18.26 & 17.15 & 1.47 & 8.58 \\
CWE-703 & 58.59 & 54.78 & 53.04 & 55.47 & 2.32 & 4.18 \\
CWE-707 & 30.75 & 29.91 & 30.96 & 30.54 & 0.45 & 1.49 \\
\bottomrule
\end{tabular}}
\end{table*}

\begin{figure*}[h]
\centering
\includegraphics[width=0.75\textwidth]{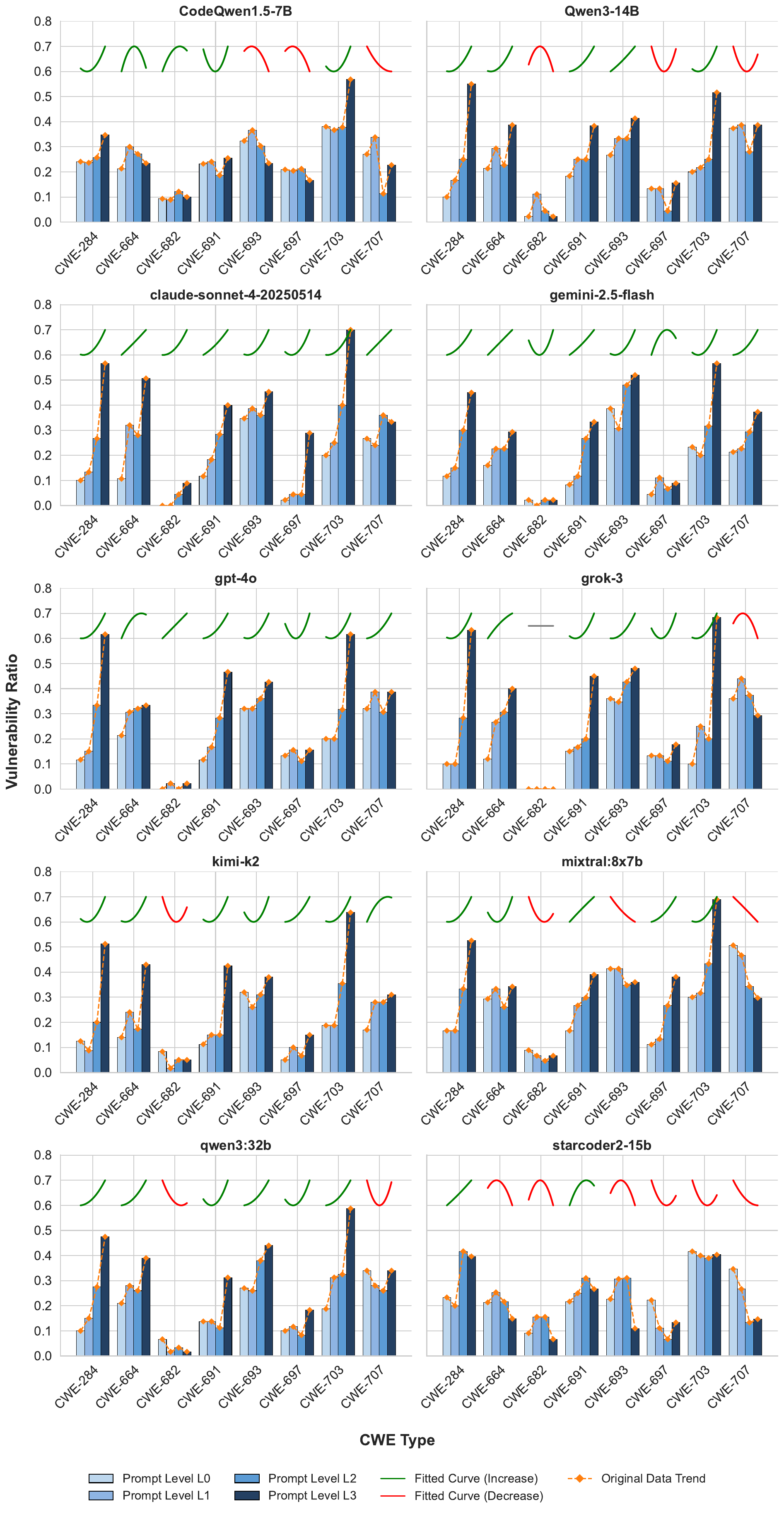}
\caption{Trend of Average Performance for Pillar CWEs across L0-L3 in Code Generation by Different Models. The green line above the bars indicates the strong correlation between code security and prompt specification.}
\label{basic-trend}
\end{figure*}

Our basic experiments first assess the direct impact of prompt normativity on code security. As shown in Figure \ref{basic-trend}, the results from all ten models across eight Pillar-CWE categories reveal a clear and universal trend: as prompt normativity decreases from L0 to L3, the vulnerability rate of the generated code significantly increases. Moreover, the trend becomes even more pronounced for larger‑parameter models that already exhibit stronger code‑generation performance, suggesting that increased model capacity amplifies the security repercussions of poorly specified prompts. This finding strongly validates our core hypothesis that the quality of requirement specification is a key determinant of code security. Furthermore, as indicated by the green trend lines above the charts, this negative correlation exhibits high consistency across all tested models, demonstrating the cross-model universality of our finding.

To provide quantitative evidence and a deeper understanding of this trend, we present detailed data in \cref{tab:vulnerability_rates_L0,tab:vulnerability_rates_L1,tab:vulnerability_rates_L2,tab:vulnerability_rates_L3} . We observe that the impact of prompt normativity is particularly pronounced for weakness categories that are intensive in logic and control flow. For example, the vulnerability rate for CWE-703 (Improper Check or Handling of Exceptional Conditions) increases substantially from 23.12\% at L0 to 58.59\% at L3, an increase of over 150\%. Similarly, the rate for CWE-284 (Improper Access Control) increases from 13.59\% to 49.84\%. These substantial increases clearly demonstrate that when requirements become logically ambiguous, an LLM's ability to implement tasks requiring rigorous logical reasoning\citep{10.24963/ijcai.2024/693}, such as correct access control and robust error handling, degrades substantially. This upward trend is visually corroborated in Figure \ref{basic-trend}, where the bar charts for CWE-703 and CWE-284 exhibit a distinct stepwise ascent for nearly all models.

Further analysis of the data reveals that not all weakness categories exhibit the same sensitivity to prompt normativity; some vulnerabilities are more resilient to declining prompt quality. As shown in \cref{tab:vulnerability_rates_L0,tab:vulnerability_rates_L1,tab:vulnerability_rates_L2,tab:vulnerability_rates_L3} , the vulnerability curve for CWE-693 (Protection Mechanism Failure) is relatively flat, with the rate increasing modestly from 31.87\% at L0 to 38.00\% at L3, a far less dramatic rise compared to logic-intensive categories. We posit that this is because the security of tasks such as cryptography or credential management depends more on the LLM's ability to recall the correct library or fixed programming patterns, rather than on complex, real-time logical reasoning. Consequently, these tasks are less susceptible to the effects of prompt ambiguity.

% Despite these observed variations in sensitivity across specific weaknesses and models (as shown in Figure \ref{basic-trend}, where, for instance, gpt-4o exhibits significantly stronger baseline security than CodeQwen1.5-7B), a universal conclusion remains evident. Without exception, all models exhibit an overall trend of increasing vulnerability rates as prompt normativity declines. This demonstrates that the impact of prompt normativity is a universal factor that influences security outcomes independently of a model's intrinsic capabilities.

Despite these observed variations in sensitivity across specific weaknesses and models (as shown in Figure \ref{basic-trend}, where, for instance, CodeQwen1.5-7B exhibits significantly stronger baseline security than Claude-Sonnet-4), a universal conclusion remains evident. Without exception, all models exhibit an overall trend of increasing vulnerability rates as prompt normativity declines. This demonstrates that the impact of prompt normativity is a universal factor that influences security outcomes independently of a model's intrinsic capabilities.

Finally, it is worth noting the non-monotonic trend exhibited by CWE-707 (Improper Neutralization), whose trend line is marked in red in Figure \ref{basic-trend}. Unlike logic-intensive weaknesses such as CWE-284, \cref{tab:vulnerability_rates_L0,tab:vulnerability_rates_L1,tab:vulnerability_rates_L2} shows that the vulnerability rate for CWE-707 is actually lower at the L2 level (27.12\%) than at L0 and L1. We posit that this difference stems from the varying security implications of the “path of least resistance” for different types of tasks. For injection-style tasks, the most direct implementation—such as string concatenation—is inherently insecure, and the clear instructions of L0/L1 prompts inadvertently guide the LLM down this “shortcut”. In contrast, the security of tasks like access control depends on logical completeness, causing their vulnerability rates to rise monotonically as prompt ambiguity increases.

\clearpage
\clearpage

\subsubsection{Cross-Language Generalization in Basic Experimental Analysis}

\begin{figure*}[h]
\centering
\includegraphics[width=0.8\textwidth]{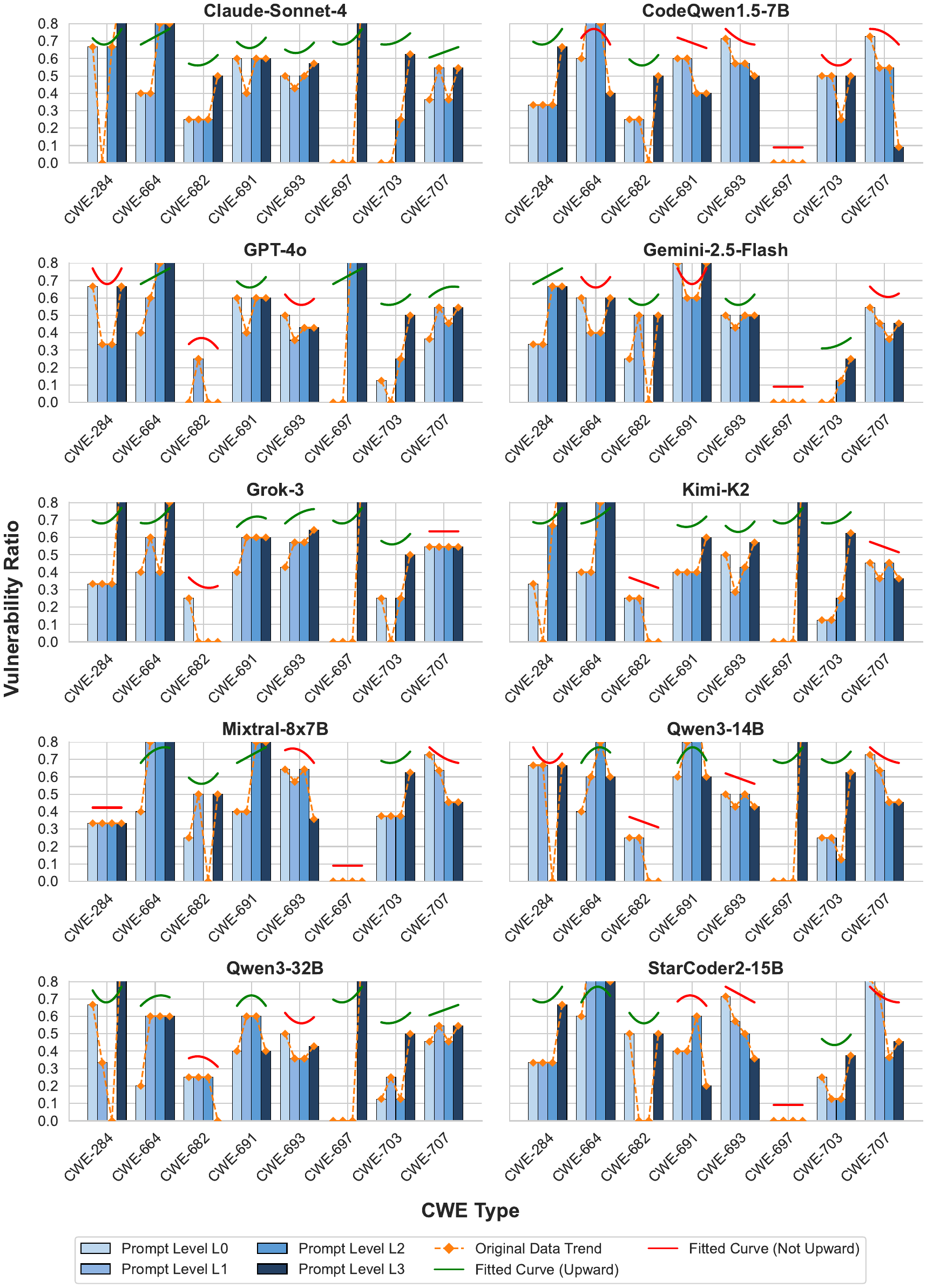}
\caption{Trend of Average Vulnerability Rates for Pillar CWEs across L0--L3 in Java Code Generation by Different Models. The trend indicators summarize the correlation between code security and prompt normativity across pillars.}
\label{fig:java_trend}
\end{figure*}

The cross-language experiments in Java reproduce the core pattern observed in Python. As shown in Figure~\ref{fig:java_trend}, across all evaluated models and eight Pillar-CWE categories, vulnerability rates increase steadily as prompt normativity decreases from L0 to L3, where L0 is the most normative setting and yields the lowest vulnerability rates, while L3 is the least normative setting and yields the highest vulnerability rates. This consistency indicates that the security impact of requirement quality persists beyond a single language setting, and remains stable under the same single-function constraint and evaluation protocol.

\begin{figure*}[h]
\centering
\includegraphics[width=0.8\textwidth]{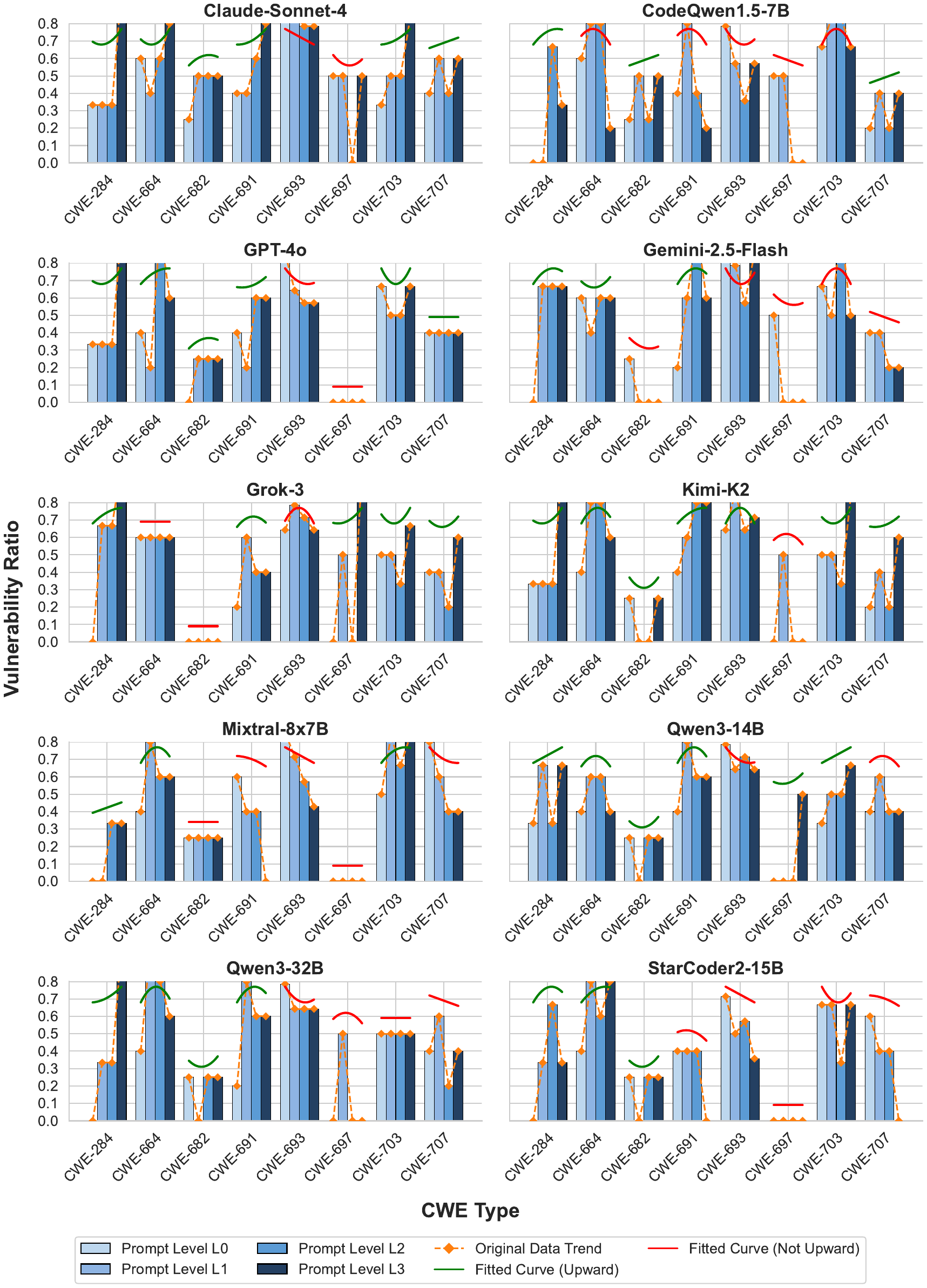}
\caption{Trend of Average Vulnerability Rates for Pillar CWEs across L0--L3 in C/C++ Code Generation by Different Models. The trend indicators summarize the correlation between code security and prompt normativity across pillars.}
\label{fig:cpp_trend}
\end{figure*}

The same relationship broadly holds in the C/C++ extension. Figure~\ref{fig:cpp_trend} shows that the L0 to L3 transition increases vulnerability rates across most pillars and models, and the rise is especially visible for pillars that depend on precise control flow, boundary checks, and error handling. However, C/C++ also exhibits a small number of model--pillar pairs with weaker monotonicity or mild non-monotonic behavior. These deviations are consistent with C/C++ introducing additional variance beyond prompt normativity. Under the single-function constraint, some tasks require auxiliary language scaffolding such as headers, type declarations, or library setup, which can amplify sensitivity to surface wording. In addition, highly normative prompts may implicitly steer models toward low-level shortcut implementations, for example fixed-size buffers, manual parsing, or C-style string manipulation, which can increase the likelihood of memory and injection related defects even at L0. Finally, a subset of C/C++ vulnerability patterns are harder to attribute from a single function without call-site context, which increases borderline cases and can occasionally weaken the pillar-level trend.

Taken together, the Java and C/C++ results provide direct evidence that prompt normativity is a language-agnostic driver of code security outcomes under a unified protocol. The dominant increase from L0 to L3 across both languages supports the generalizability of our conclusion that more normative requirement descriptions systematically reduce vulnerability risks even when prompts contain no security directives, while the limited exceptions in C/C++ highlight language-specific constraints that motivate future extensions with minimal scaffolding and context-aware checks.

\clearpage
\clearpage

\clearpage
\clearpage

\subsubsection{Perturbation-Based Experimental Analysis}    

To validate the robustness of our core finding that prompt normativity is strongly correlated with code security, we performed supplementary experiments on the perturbation augmented version of the base dataset. Specifically, we repeated the full evaluation using two semantically equivalent but subtly altered prompt variants designed during dataset construction:  (1) lexical and syntactic perturbations (v1) and (2) structural and information flow perturbations (v2). We hypothesize that if our baseline results are truly robust, the negative correlation observed between prompt normativity and code security will persist in the perturbed dataset.

\begin{figure*}[h]
    \centering
    \includegraphics[width=\textwidth]{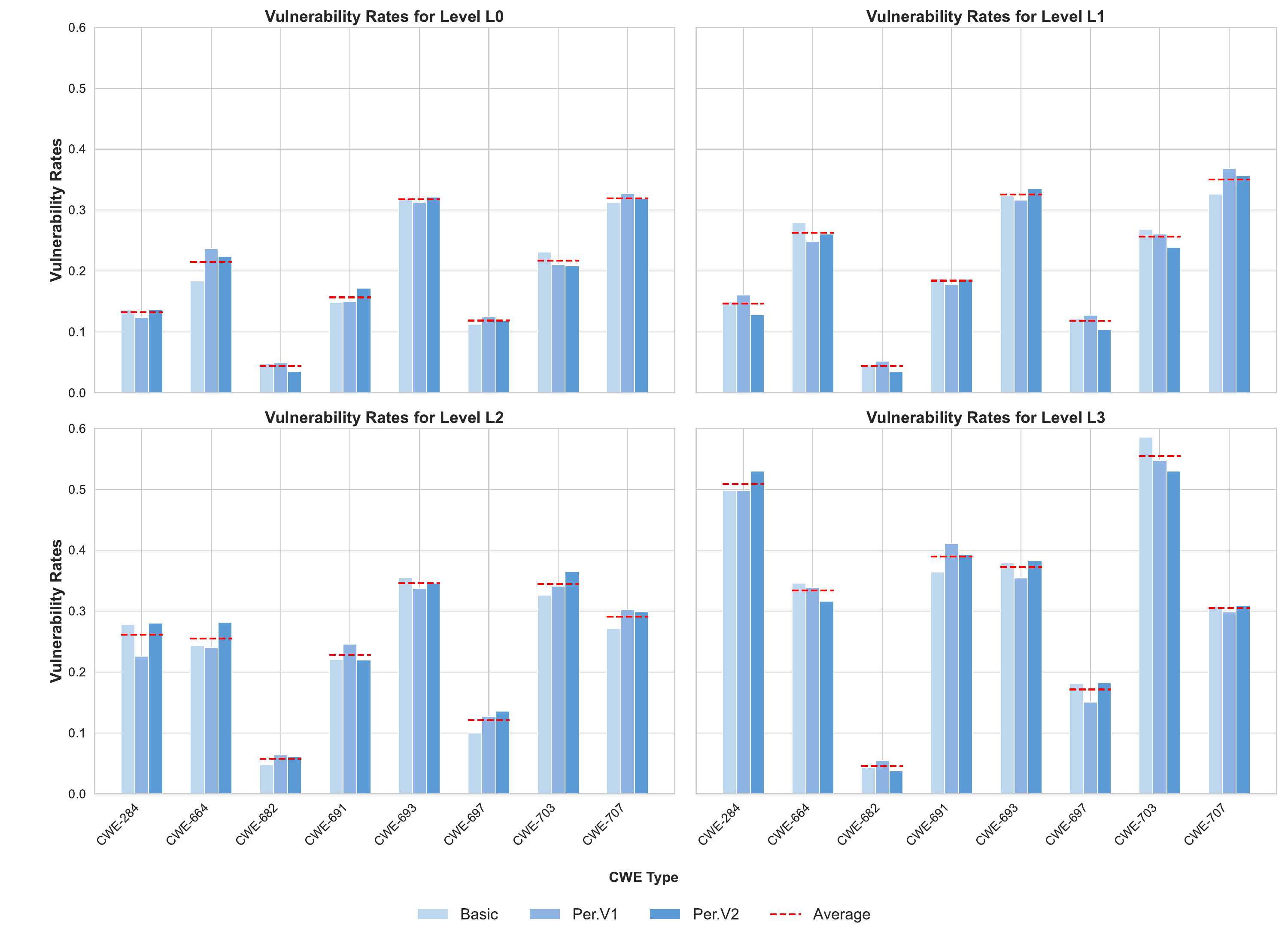}
    \caption{Comparative Analysis of Vulnerability Rates between the Perturbation Experiment and the Baseline}
    \label{fig:basic_syn_v12}
\end{figure*}

First, we evaluated the impact of the v1 lexical and syntactic perturbations. Figure \ref{fig:basic_syn_v12} shows the trends of the vulnerability rate for each Pillar-CWE in different prompt levels in this perturbed version. By comparing these results with the trends of the baseline experiment, we found a high degree of consistency. For any given Pillar CWE, the trends observed in both the baseline and the perturbation v1 experiments align closely across the prompt levels. This indicates that our core observed trend is not triggered by specific keywords, but is robustly replicated across semantically equivalent sets of prompts at all levels.

% Next, we evaluated the impact of the more challenging structural and informational perturbations of v2-structure, with the results presented in Table \ref{tab:basic_syn_v12}. Although this perturbation altered the narrative structure of the requirements, the core trend we observed remained robust. The results for Perturbation v2 in Table \ref{tab:basic_syn_v12} once again replicated the systematic downward pattern from the baseline experiment: the security of the code generated by all models still decreased as prompt regularity declined. This phenomenon demonstrates that our conclusion holds true even under variations in the order of information presentation.

Next, we evaluated the impact of the more challenging v2 structural and informational perturbations. The results once again replicated the systematic downward pattern from the baseline experiment: the security of the code generated by all models still decreased as prompt normativity declined. This phenomenon further demonstrates that our conclusion holds true even under variations in the order of information presentation. The vulnerability rates of the three experiments (Basic, v1, and v2) are remarkably similar for the same Pillar-CWE. This strong consistency across different prompt levels further confirms the robustness of the vulnerability rate trends. %Therefore, we confidently assert that structural changes and word order variations have a minimal impact on the overall trend of vulnerability rates.
%Thus, the convergent patterns across Basic, v1, and v2 validate our main thesis: decreasing prompt normativity consistently elevates vulnerability rates, independent of semantically preserving lexical–syntactic perturbations (v1) and structural/informational-flow reordering (v2).

%
As shown in figure \ref{fig:basic_syn_v12}, the vulnerability rates of the three experiments are similar for the same Pillar-CWE. Meanwhile, based on the preceding analysis of the 'basic' experiment and the 'Perturbation v1' and 'v2' experiments, the trends in vulnerability rates across different prompt levels are consistent for the same Pillar-CWE. This indicates that structural changes and word order variations have a minimal impact on the trend of the vulnerability rate, thereby demonstrating the robustness of the results from the basic experiment.

\subsection{Model generalization trend}
The baseline experiments confirm our core hypothesis: the data clearly demonstrate a causal link between prompt normativity and the security of LLM‐generated code. As prompt normativity, defined as the quality of functional requirements specification, deteriorates, the model is forced to make more guesses and autonomous decisions under conditions of incomplete information and logical conflict. This, in turn, greatly increases the likelihood that it will choose the “simplest” rather than the “safest” path, leading to the generation of insecure code. These findings offer crucial empirical evidence for understanding and mitigating the security risks of LLMs in real‐world applications.

By extending our evaluation to the perturbation‐augmented dataset, we observe that within-group differences across the Basic and perturbed versions are minimal, while between-level differences across L0–L3 remain highly significant. This sharp contrast further validates the robustness of our results. It shows unambiguously that the observed decline in security rate is driven by prompt normativity itself—and not by any accidental choice of wording or sentence structure at each level—thus reinforcing the stability and generalizability of our baseline conclusions and solidly supporting the claim that prompt normativity is a key determinant of LLM code‐generation security.

\section{Security Potential Validation Experiments}
To explore which prompt normativity leads to safer code generation, we selected two practical prompting strategies commonly used by everyday developers and evaluated their effectiveness in large‐scale experiments. We deliberately excluded multi‐turn dialogue optimization, RAG, and agent‐driven approaches, since these methods can amplify LLM hallucination effects—thereby compromising evaluation accuracy—and introduce additional variables that would hinder the rigorous validation of our core conclusions.

\subsection{Optimization of Generation Methods}
\subsubsection{Chain-of-Thought}
Compared to traditional prompting methods, chain-of-thought (CoT) \citep{10.5555/3600270.3602070} prompting involves a series of interconnected queries, with each prompt building upon the previous one, continuously refining the analysis and providing richer context. For example, when generating a Python function for user credential verification, the first prompt might request the function to return a boolean indicating the validity of a given username and password. The initial response would focus on enhancing security by using hashed password comparisons rather than plain-text storage. This illustrates a simplified form of CoT's iterative reasoning process, which can be mathematically expressed as:
\[
P_{\text{CoT}} = \{S, (x_1, e_1), \dots, (x_n, e_n)\},
\]
where \(P_{\text{CoT}}\) represents the prompt used for \textbf{CoT reasoning}, and \(e_i\) represents the example rationale for each step.  As the conversation progresses, subsequent prompts would ask how to protect inputs and prevent vulnerabilities like timing attacks, with the LLM suggesting techniques such as constant-time comparisons and rate-limiting mechanisms to defend against brute-force attacks. This process is expressed as:

\[
p(A | P_{\text{CoT}}, Q, R) = p(A | P_{\text{CoT}}, Q) \cdot p(R | P_{\text{CoT}}, Q),
\]
where \(p(A | P_{\text{CoT}}, Q)\) represents the probability of generating the answer \(A\) given the prompt \(P_{\text{CoT}}\) and a query \(Q\), while \(p(R | P_{\text{CoT}}, Q)\) models the probability of generating the rationale \(R\). Each additional prompt further refines the function, focusing on secure password storage and implementing multi-factor authentication. By adopting this step-by-step reasoning approach, the code quality generated by the CoT method improves. The iterative process reflects how complex, security-focused code evolves through collaborative discussion, with each step contributing to the final, logically consistent, and secure solution.
 
In contrast to traditional methods, which may lack logical coherence in their generation process, CoT reasoning introduces a step-by-step reasoning framework that ensures each step in the code generation process is logically connected. It minimizes the risk of missing crucial security checks or logical inconsistencies. By making explicit the rationale behind each generation step, CoT improves the reliability and correctness of the code.

\subsubsection{Regenerate Act}

Our optimization phase's Self-Correction approach draws inspiration from the SELF-REFINE framework\citep{10.5555/3666122.3668141}. The core idea behind SELF-REFINE is that, much like humans, Large Language Models can improve their initial outputs through iterative feedback and refinement, without requiring additional training data or reinforcement learning. Inspired by this, our “Regenerate Act” method is designed to optimize code from a security perspective. This method proceeds in two steps: first, the LLM generates an initial code; then, it leverages its internal knowledge base and experience to simulate the role of a security expert, performing vulnerability assessment on its own generated code to identify potential security vulnerabilities and attempt to fix them based on known security patterns and best practices learned from the model's training data.

\begin{figure}[h!]
    \centering
    \includegraphics[width=\textwidth]{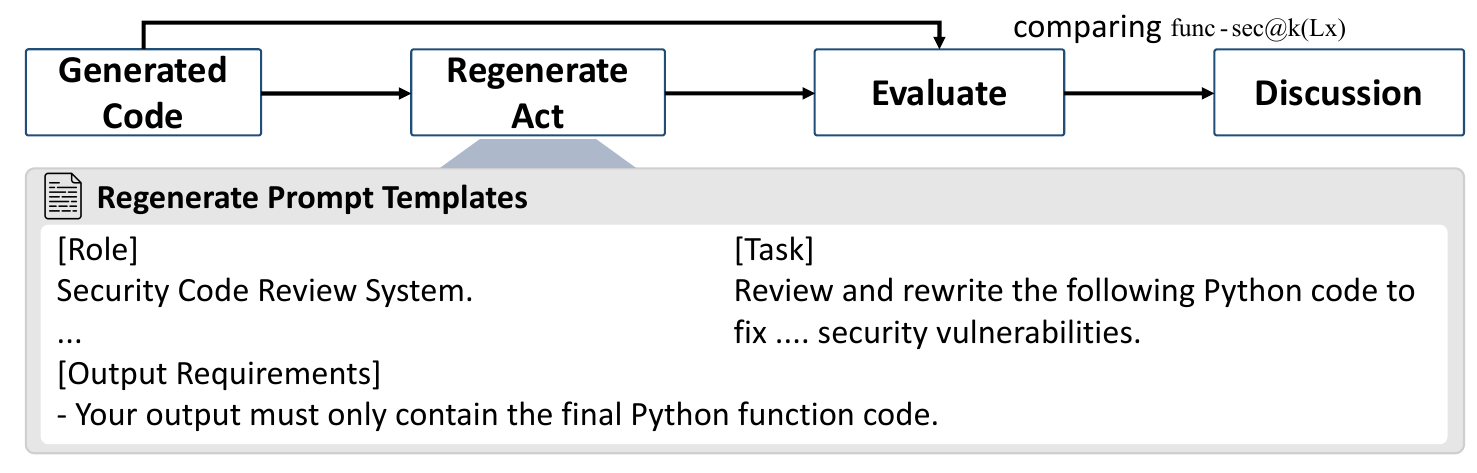}
    \caption{Regenerate Act Process and Template Overview}
    \label{fig:regenerate_flowchart}
\end{figure}

Through a template, the LLM is configured to act as a code security expert. It assesses vulnerabilities in the code from both the code and business perspectives, providing a fully optimized version of the code after each evaluation. By comparing the vulnerability rate with that of the previously generated original code, the effectiveness of the self-checking process is validated, exploring whether it has the capability to generate secure code, as shown in Figure~\ref{fig:regenerate_flowchart}.

\subsection{Security Mitigation with Optimization Methods}

Table~\ref{tab:vulnerability_rates} presents the vulnerability rates for Pillar-CWEs across different prompt levels under the Basic, CoT, and RE-ACT optimization methods. Both COT and RE-ACT optimization methods significantly reduce the vulnerability rates, especially in the case of more complex vulnerability types. Through the COT mechanism and the Self-Review mechanism, the security quality of code generation is substantially improved. However, the changes in vulnerability rates across different CWE tasks are not consistent, reflecting the impact of varying task difficulty on the vulnerability rates.

\input{attachments/tables/vulnerability_rates}

\begin{figure*}[h]
\centering
\includegraphics[width=0.9\textwidth]{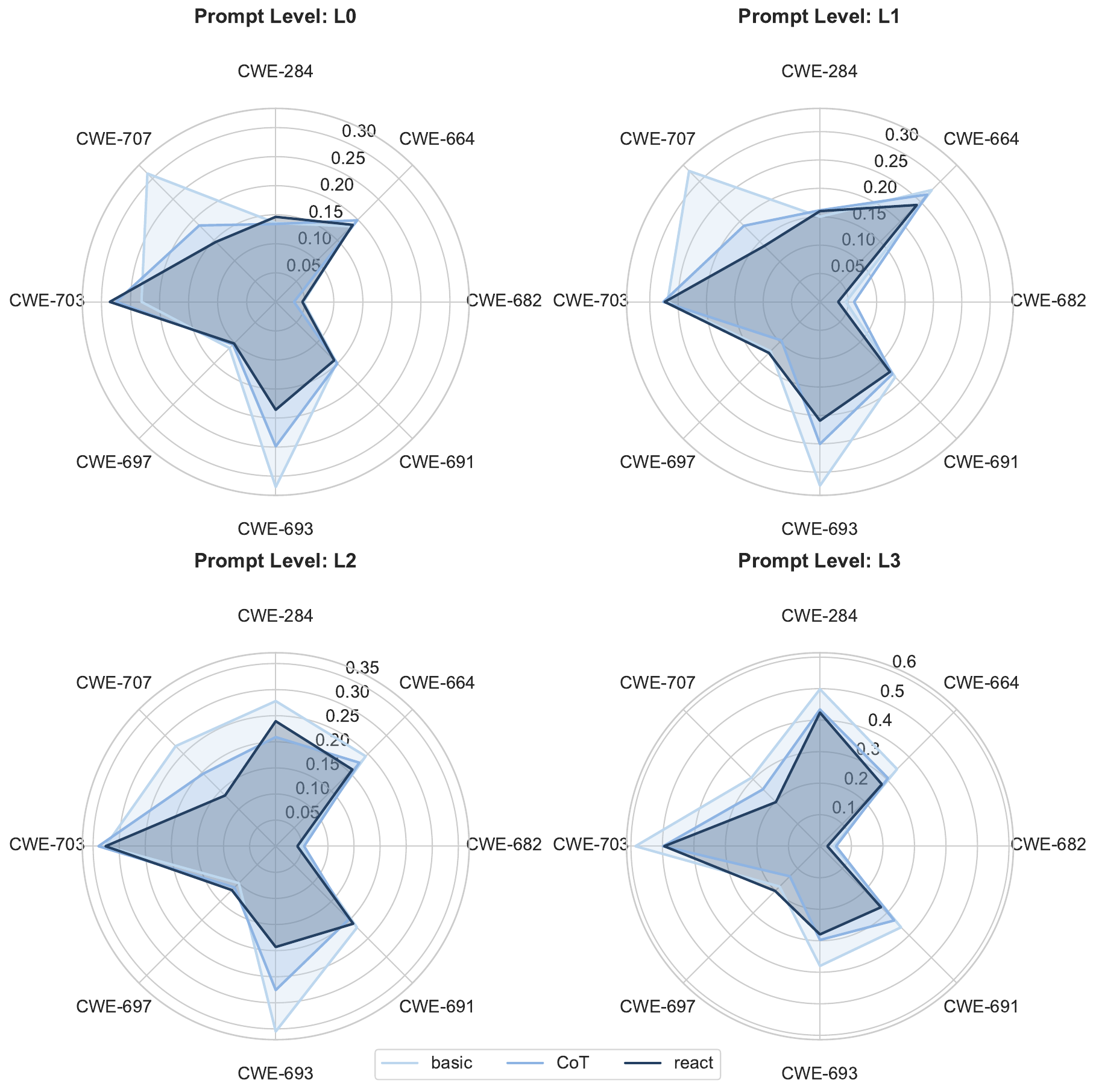}
\caption{Comparison of CoT and Re-Act Optimization Methods for Code Generation Security and Vulnerability Rate Improvement at Various Prompt Levels.}
\label{basic_cot_reatc_compared}
\end{figure*}

Figure~\ref{basic_cot_reatc_compared} displays the radar distribution of the vulnerability rate changes for different CWE vulnerabilities under various optimization methods across four specification levels (L0 to L3). From the figure, it is evident that as the specification level increases, both COT and RE-ACT methods lead to a reduction in vulnerability rates across different CWE tasks. Notably, in more complex tasks like CWE-284 and CWE-693, both COT and RE-ACT show significant reductions in vulnerability rates at the L3 level. The radar plot shows that COT exhibits a more significant reduction in vulnerability rates compared to the basic method  for most CWE tasks, especially in high-difficulty tasks such as CWE-284 and CWE-693. This suggests that the COT optimization method can further reduce the vulnerability rate at higher specification levels. While RE-ACT (in gray) does not always outperform COT, it demonstrates a prominent protective capability, particularly at L2 and L3 levels in complex tasks like CWE-693.

The COT optimization method enhances the security of the code generation process by leveraging the COT mechanism. This approach enables the model to reason step-by-step, fostering explicit security considerations at each stage of code generation, and thereby reducing potential vulnerabilities. For low-difficulty CWE tasks, such as CWE-682, where the base vulnerability rate is already low (L0: 4.79\%), COT optimization yields only minor reductions, suggesting that for such straightforward vulnerabilities, the models inherently possess strong secure generation capabilities. Conversely, in more complex tasks like CWE-284 and CWE-693, the advantages of COT optimization become strikingly pronounced.
Specifically, for CWE-284, COT demonstrably reduces the vulnerability rate from 49.84\% to 43.41\% at L3 and from 27.81\% to 20.91\% at L2. A marginal decrease from 13.59\% to 13.41\% is also observed at L0. While a slight increase from 15.00\% to 16.14\% is noted at L1, this isolated instance of modest fluctuation does not detract from the overall positive impact of COT. This trend signifies that COT is particularly effective in reducing vulnerability rates at higher ambiguity levels , where the model's inferential burden is substantially greater, leading to the elimination of more potential vulnerabilities. The COT mechanism thus significantly enhances the model's ability to handle complex vulnerabilities by intensifying its focus on security during the multi-step reasoning process.

\begin{figure}[h!]
    \centering
    \includegraphics[width=\textwidth]{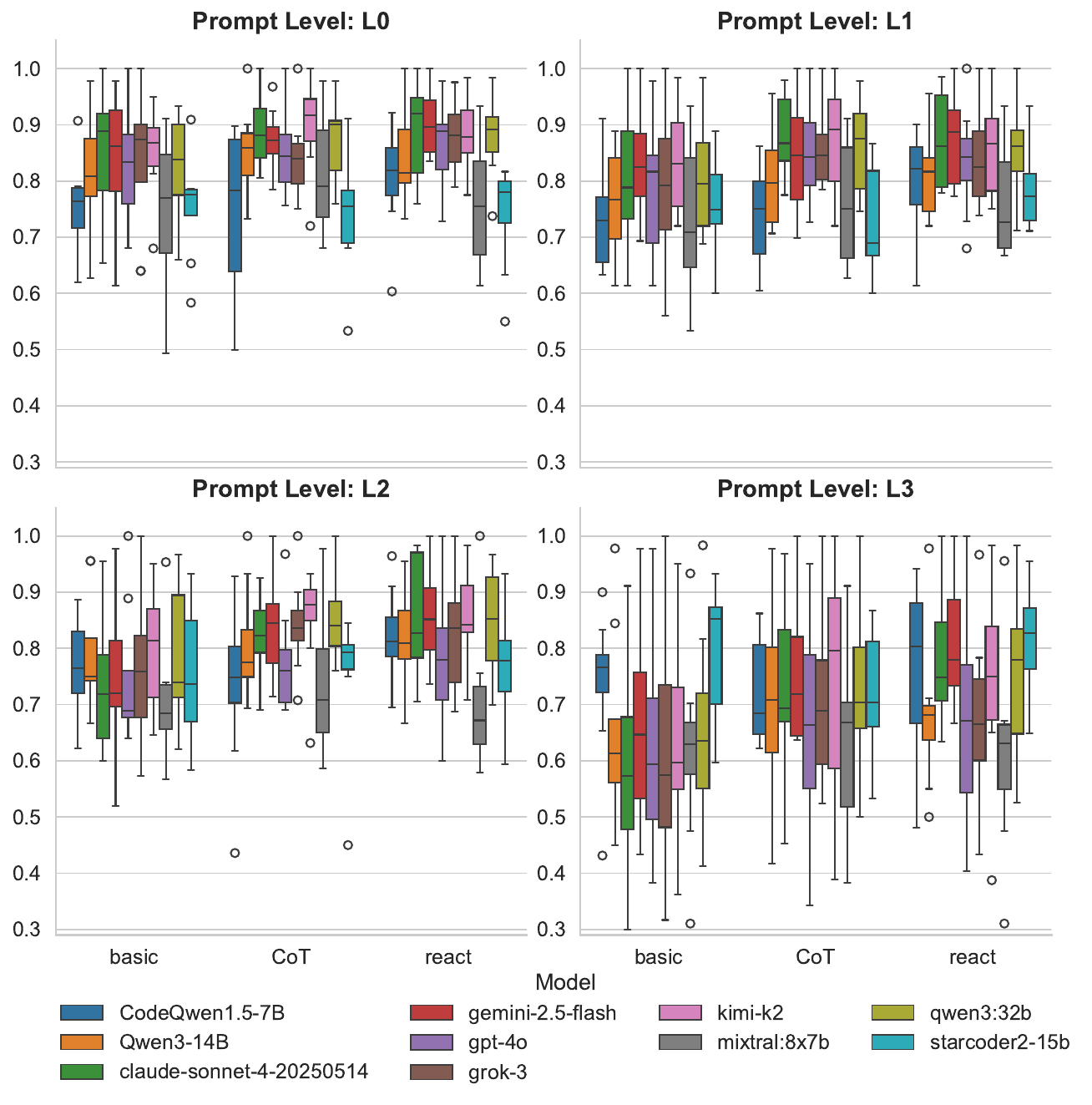}
    \caption{Performance Comparison of Models under Basic, COT, and RE-ACT Optimization Methods across Different Prompt Levels}
    \label{basic_cot_reatc_box_line}
\end{figure}

Figure~\ref{basic_cot_reatc_box_line} presents the distribution of different models under various optimization methods, illustrating how the level of prompt specification corresponds to the degree of dispersion in performance. At the L0 level, where prompts are fully standardized, the boxplots remain compact with short boxes, limited whiskers, and only a few outliers, indicating that performance is relatively concentrated and stable. Models such as CodeQwen1.5-7B and Qwen3-14B both show tight distributions at this level across methods, confirming that complete standardization effectively constrains variability. At the L1 level, where prompts are only partially standardized, the dispersion becomes larger as boxes and whiskers lengthen and more outliers appear. In this setting, Gemini-2.5-flash exhibits a wide performance spread under the Basic method, but its distribution becomes more compact with CoT optimization, suggesting that CoT helps mitigate instability introduced by reduced prompt regularity. At the L2 level, with unstandardized prompts, the variability is more pronounced, reflected in wider boxes, longer whiskers, and frequent outliers. Mixtral-8x7B shows a markedly broader distribution under the Basic method, but CoT optimization narrows the interquartile range, reducing fluctuation compared to the unoptimized case. At the L3 level, where prompts are highly unstandardized, the distributions are the most dispersed with long boxes, extended whiskers, and numerous outliers, highlighting strong instability. In this context, the advantage of RE-ACT optimization becomes more evident, as models such as GPT-4o and Grok-3 exhibit smaller fluctuation ranges and fewer outliers compared with Basic or CoT. Moreover, Mixtral-8x7B demonstrates a visibly more compact distribution under RE-ACT at L3, showing that the self-review mechanism can partially counteract instability caused by highly unstandardized prompts. The results show that while models have safe code generation capability, standardized prompts are essential for stability, as less standardized prompts increase performance dispersion.

The RE-ACT optimization method significantly enhances secure code generation through its robust Self-Review Mechanism. This enables the model to perform real-time vulnerability checks and fixes via self-feedback during the generation process, proving particularly effective for high-complexity tasks. For instance, in the CWE-284 task, RE-ACT reduces the vulnerability rate from 27.81\% to 23.95\% at L2 and from 49.84\% to 42.41\% at L3. At L0, the vulnerability rate for CWE-284 shows a marginal increase from 13.59\% to 14.64\% under RE-ACT optimization, which contrasts with the general downward trend at higher ambiguity levels. This specific observation, however, does not diminish RE-ACT's overall positive impact, especially as its strengths become evident in more ambiguous scenarios. Although RE-ACT does not consistently outperform COT across all tasks, its self-review mechanism markedly improves system security in specific scenarios. Notably, in tasks with intrinsically higher vulnerability rates, such as CWE-693, RE-ACT achieves substantial reductions at L2 and L3. This underscores RE-ACT's crucial role in complex vulnerability protection, utilizing self-feedback and dynamic repair to refine output for security post-generation.

\section{Additional Analyses and Diagnostics}

This section provides additional analyses and diagnostics that complement the main results.
Across all four subsections, we adopt a unified LLM-based judging protocol to ensure consistent evaluation.
Specifically, we use three judge models (DeepSeek-V3.1, GPT-4o, and Grok-3) and aggregate their decisions via majority voting.
We first report functional correctness across prompt normativity levels (Section~\ref{func}), then re-estimate vulnerability rates under the same voting protocol (Section~\ref{vote_vul}).
Next, we examine robustness to decoding temperature while keeping the evaluation protocol fixed (Section~\ref{temp}).
Finally, we provide model-level diagnostics on factors such as scale and openness under the same evaluation setting (Section~\ref{model_diag}).

\subsection{Functional Correctness Assessment}\label{func}

To contextualize the security findings, we report functional correctness as an additional diagnostic for LLM-generated code across prompt normativity levels.
We evaluate whether each generated solution satisfies the stated functional requirements using three LLM judges with majority voting, and aggregate results over all prompting variants including basic prompting, synonym expansion, CoT, and ReAct.

\begin{figure*}[h]
    \centering
    \includegraphics[width=\textwidth]{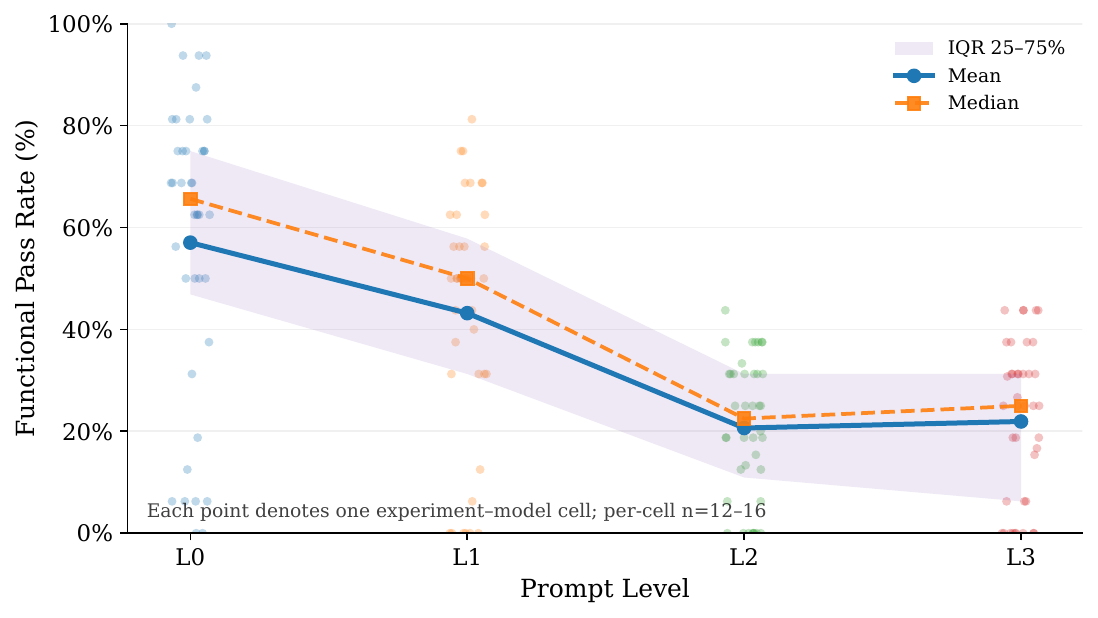}
    \caption{Functional pass rate across prompt normativity levels aggregated over all prompting variants and models. Points show per cell pass rates under majority voting.}
    \label{fig:func_pass}
\end{figure*}

Figure~\ref{fig:func_pass} summarizes functional pass rates from L0 to L3.
Each dot corresponds to one experiment--model cell and reflects the voting-based pass rate on a fixed subset of tasks, while the shaded band indicates the interquartile range across cells.
Overall, functional correctness degrades systematically as prompt normativity weakens: both the mean and median pass rates decline markedly from L0/L1 to L2/L3, and dispersion increases under weaker prompts, indicating larger variability across models and prompting settings.

This pattern aligns with the role of normativity as an external constraint.
Highly normative prompts more explicitly specify required behaviors and edge conditions, reducing underspecification and making it easier for models to produce solutions that meet functional requirements.
In contrast, less normative prompts provide fewer explicit constraints, increasing the likelihood of partial solutions or misalignment with the intended task.
As a consequence, prompt normativity affects not only vulnerability outcomes but also the reliability of task completion.

\subsection{Vulnerability Assessment under Majority Voting}\label{vote_vul}

Building on the functional correctness analysis in Section~\ref{func}, we next examine vulnerability rates under the same majority-voting evaluation protocol.

\begin{figure*}[h]
    \centering
    \includegraphics[width=\textwidth]{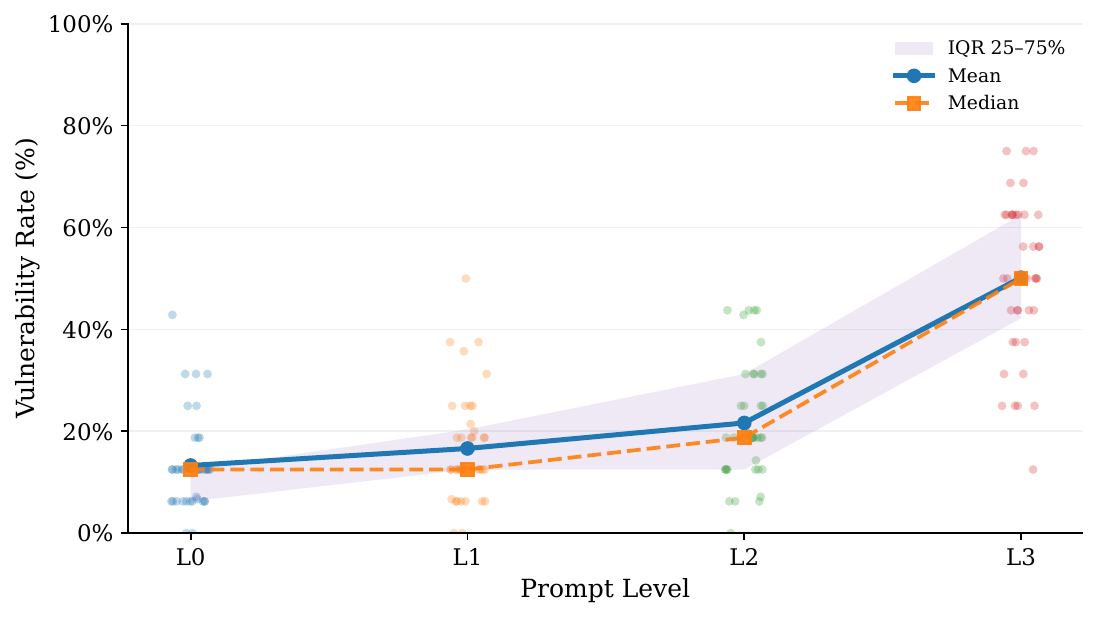}
    \caption{Vulnerability rate across prompt normativity levels under majority voting, aggregated over all settings.}
    \label{fig:vul_vote}
\end{figure*}

Figure~\ref{fig:vul_vote} reports vulnerability rates from L0 to L3 under majority voting.
Across all settings, vulnerability rates remain relatively low and stable under highly normative prompts (L0–L1), but increase steadily as prompt normativity weakens.
In particular, a pronounced rise is observed from L2 to L3, where both the mean and median vulnerability rates increase sharply, accompanied by a substantial expansion of the interquartile range.

These results indicate that the security trends reported in the main analysis are robust to the use of majority voting.
The observed increase in vulnerability under less normative prompts is not driven by individual judge variability, but persists when aggregating judgments across multiple independent evaluators.
Together with the functional correctness results in Section~\ref{func}, this analysis shows that reduced prompt normativity simultaneously degrades task reliability and amplifies security risks, reinforcing prompt normativity as a key factor shaping LLM code generation behavior.

\subsection{Robustness to Decoding Temperature}\label{temp}

We further examine whether the relationships between prompt normativity, functional correctness, and vulnerability rates are sensitive to decoding hyperparameters.
Specifically, we vary the sampling temperature during code generation while keeping the evaluation protocol unchanged, using the same three LLM judges and majority voting.
We report results for a closed-source model (GPT-4o) and an open-source model (Qwen3-32B) to cover both model types.

\begin{figure*}[ht]
    \centering

    \begin{subfigure}[t]{\textwidth}
        \centering
        \includegraphics[width=\textwidth]{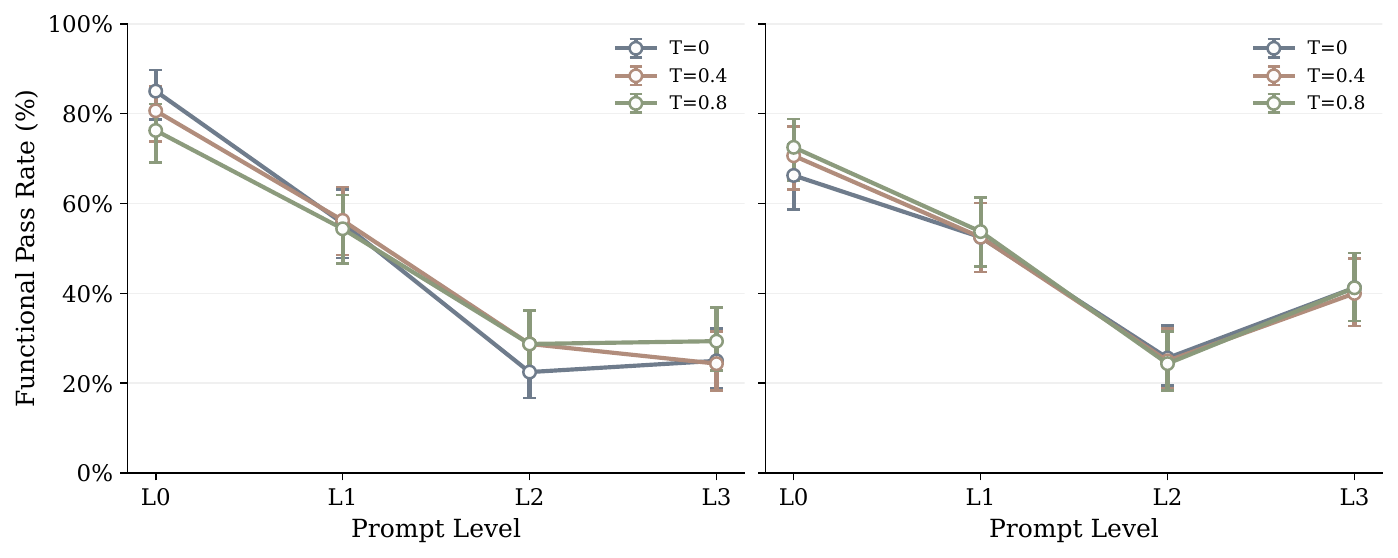}
        \caption{}
        \label{fig:temp_func}
    \end{subfigure}

    \vspace{0.8em}

    \begin{subfigure}[t]{\textwidth}
        \centering
        \includegraphics[width=\textwidth]{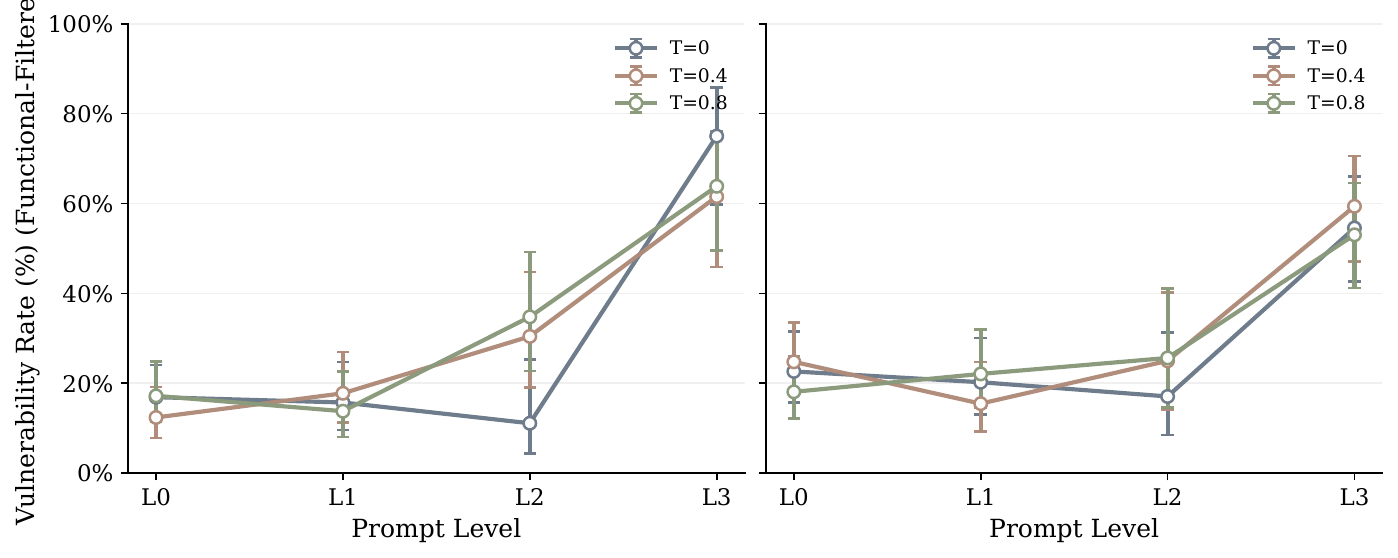}
        \caption{}
        \label{fig:temp_vul}
    \end{subfigure}

    \caption{Temperature robustness across prompt normativity levels for GPT-4o and Qwen3-32B under a fixed majority-voting evaluation protocol. The top panel reports functional pass rates, and the bottom panel reports vulnerability rates computed on the subset of functionally correct solutions.}
    \label{fig:temp_robust}
\end{figure*}

\begin{figure*}[t]
    \centering

    \begin{subfigure}[t]{0.49\textwidth}
        \centering
        \includegraphics[width=\textwidth]{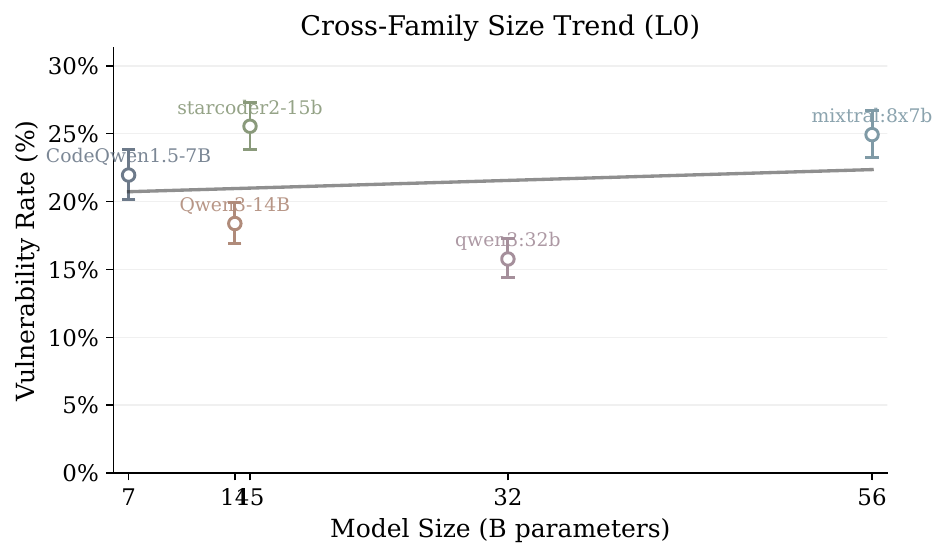}
        \caption{}
        \label{fig:size_L0}
    \end{subfigure}
    \hfill
    \begin{subfigure}[t]{0.49\textwidth}
        \centering
        \includegraphics[width=\textwidth]{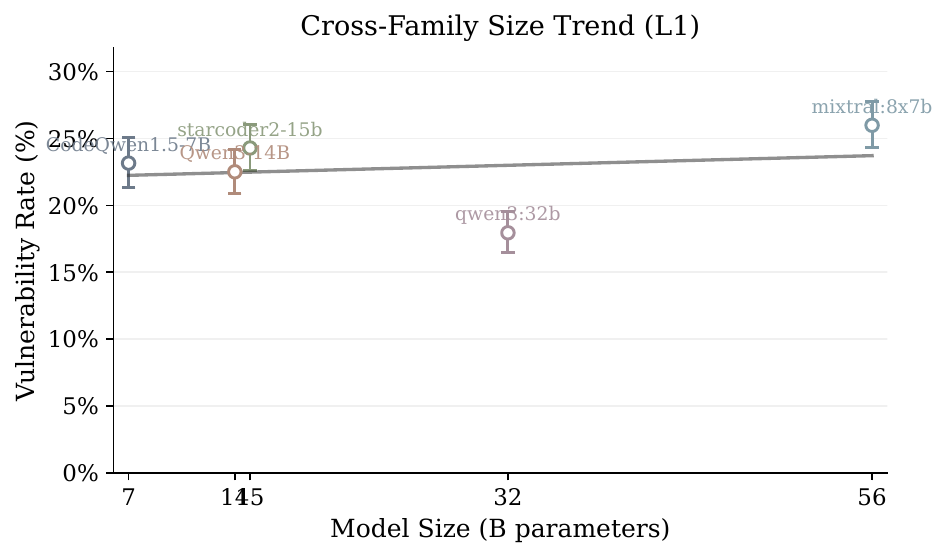}
        \caption{}
        \label{fig:size_L1}
    \end{subfigure}

    \vspace{0.6em}

    \begin{subfigure}[t]{0.49\textwidth}
        \centering
        \includegraphics[width=\textwidth]{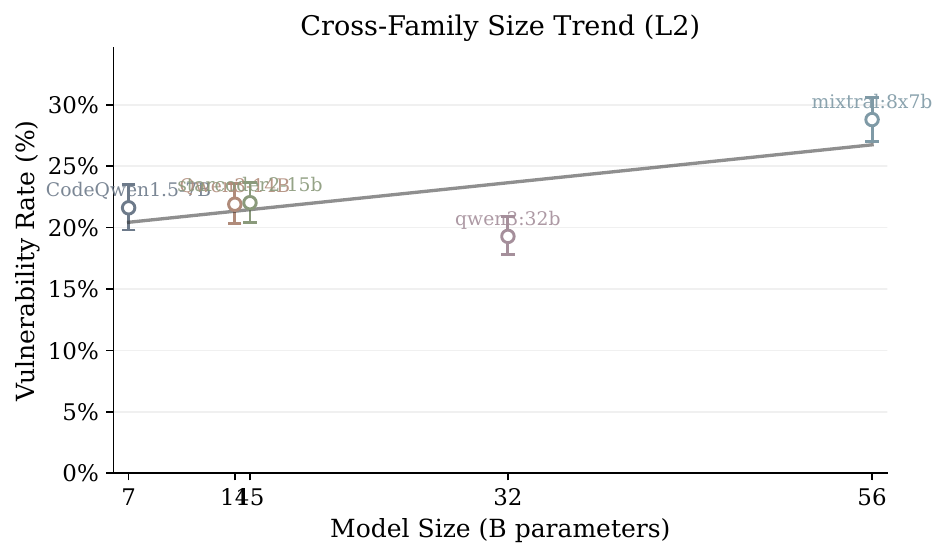}
        \caption{}
        \label{fig:size_L2}
    \end{subfigure}
    \hfill
    \begin{subfigure}[t]{0.49\textwidth}
        \centering
        \includegraphics[width=\textwidth]{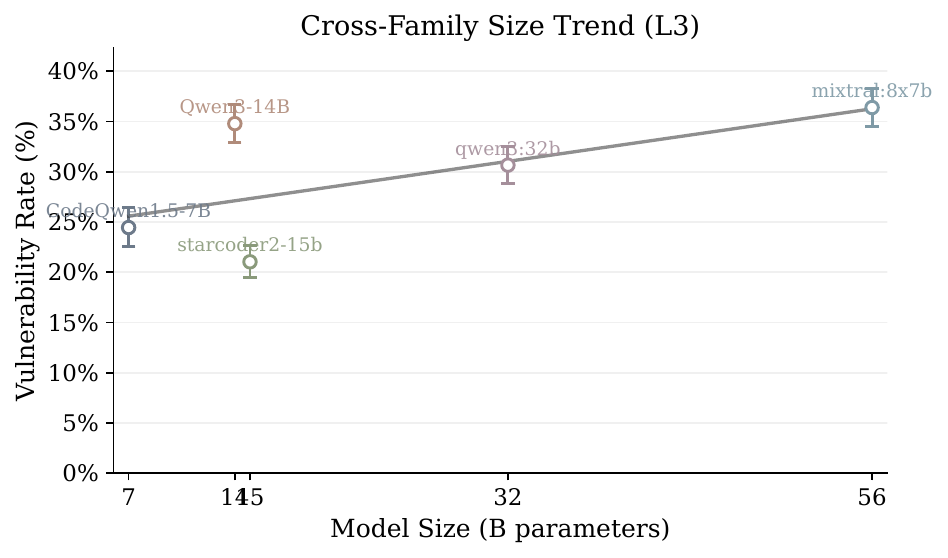}
        \caption{}
        \label{fig:size_L3}
    \end{subfigure}

    \caption{Cross-family size trend under fixed prompt normativity levels using open-source models. Each panel corresponds to one prompt level from L0 to L3 and plots vulnerability rate against parameter count. The fitted line summarizes the overall trend within the panel.}
    \label{fig:cross_family_size}
\end{figure*}

\begin{figure*}[!t]
    \centering

    \includegraphics[width=\textwidth]{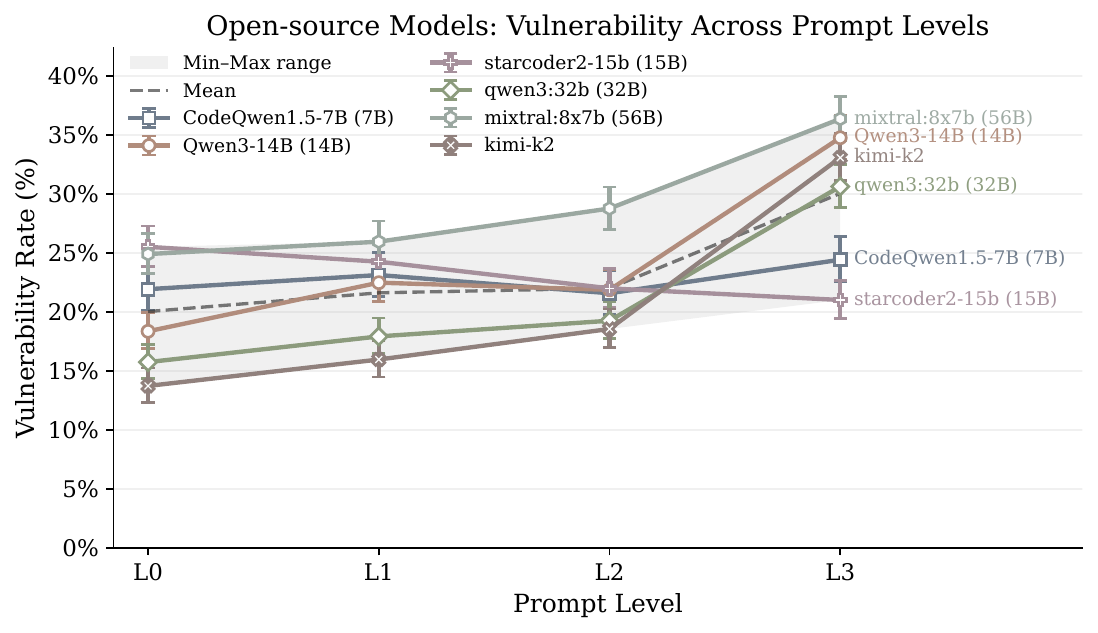}

    \vspace{0.8em}

    \includegraphics[width=\textwidth]{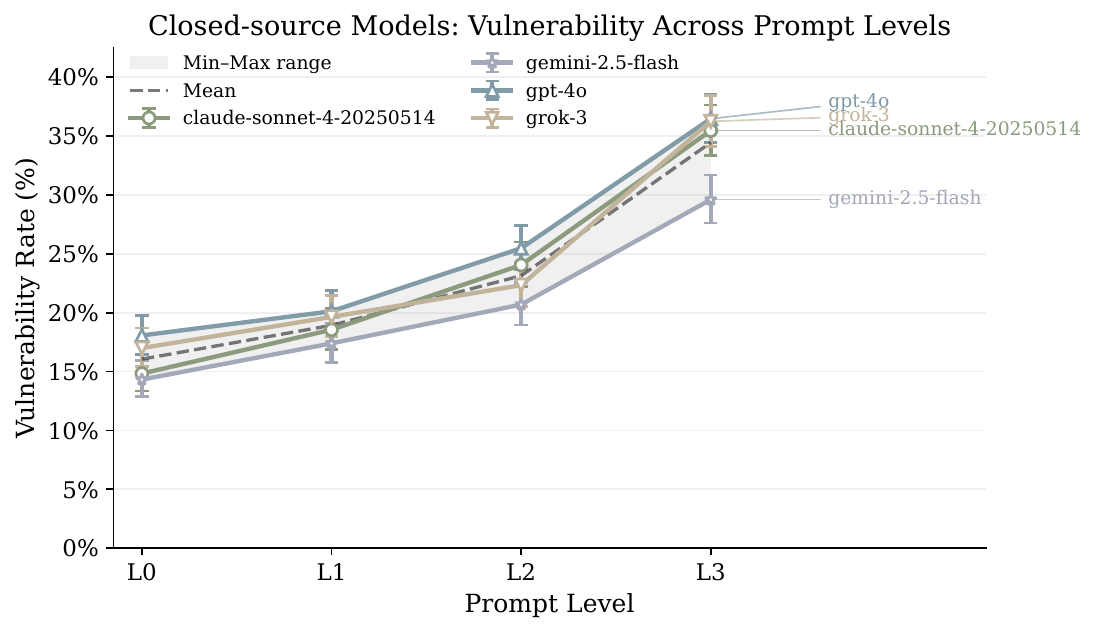}

    \caption{Vulnerability rates across prompt normativity levels for open-source and closed-source models. The top panel reports open-source models and the bottom panel reports closed-source models. For each panel, the shaded region indicates the min--max range across models and the dashed line denotes the mean trend.}
    \label{fig:model_group_vul}
\end{figure*}

%\begin{figure*}[t]
%    \centering
%
%    \begin{subfigure}[t]{\textwidth}
%        \centering
%        \includegraphics[width=\textwidth]{attachments/images/fig_model_diag_vul_by_prompt_open.pdf}
%        \caption{}
%        \label{fig:model_group_open}
%    \end{subfigure}
%
%    \vspace{0.8em}
%
%    \begin{subfigure}[t]{\textwidth}
%        \centering
%        \includegraphics[width=\textwidth]{attachments/images/fig_model_diag_vul_by_prompt_closed.pdf}
%        \caption{}
%        \label{fig:model_group_closed}
%    \end{subfigure}
%
%    \caption{Vulnerability rates across prompt normativity levels for open-source and closed-source models. The top panel reports open-source models and the bottom panel reports closed-source models. For each panel, the shaded region indicates the min--max range across models and the dashed line denotes the mean trend.}
%    \label{fig:model_group_vul}
%\end{figure*}

Figure~\ref{fig:temp_robust} summarizes temperature robustness from two complementary perspectives.
Figure~\ref{fig:temp_robust}\subref{fig:temp_func} shows functional pass rates across normativity levels.
Across temperatures, both models exhibit a consistent decline in functional correctness as prompts become less normative, with the most notable drop occurring from L1 to L2.
While temperature affects the absolute pass rate to some extent, the overall ordering across prompt levels remains stable.

Figure~\ref{fig:temp_robust}\subref{fig:temp_vul} reports vulnerability rates for both models. Vulnerability rates increase as prompt normativity weakens, and the rise becomes pronounced under L3.
Temperature mainly modulates the magnitude of vulnerability rates, but does not alter the qualitative trend that weaker prompts are associated with higher security risk.
Overall, these results indicate that the normativity-driven patterns reported in Section~5 and Sections~\ref{func}--\ref{vote_vul} are not artifacts of a particular decoding temperature.

\subsection{Model-Level Diagnostics}\label{model_diag}

We further provide model-level diagnostics to complement the normativity-driven findings in Section~5 and Sections~\ref{func}--\ref{temp}.
This analysis focuses on two aspects: how model scale relates to vulnerability rates when comparing models across families, and how vulnerability rates vary across open-source and closed-source model groups under different prompt normativity levels.

\paragraph{Cross-family size trends under different normativity levels.}
To isolate the association between model scale and vulnerability under a fixed normativity condition, Figure~\ref{fig:cross_family_size} plots vulnerability rate against parameter count for open-source models, separately for each prompt level.
Under highly normative prompts (L0--L1), the fitted trend is weak and the points are widely dispersed, indicating that explicit task constraints largely dominate security outcomes and diminish the observable effect of scale.

As normativity decreases (L2--L3), the positive association becomes more pronounced: larger models tend to exhibit higher vulnerability rates, and the fitted trend line steepens, particularly at L3.
These results are consistent with the interpretation that increased generative capacity can amplify security risks when external constraints are insufficient.
Importantly, this pattern should be interpreted as a conditional association rather than a causal effect of size, since cross-family comparisons inevitably confound scale with differences in model architecture, training data, and alignment procedures.

\paragraph{Open-source versus closed-source models across prompt levels.}
Figure~\ref{fig:model_group_vul} contrasts vulnerability trajectories across prompt levels for open-source and closed-source models, respectively.
Both groups exhibit a consistent increase in vulnerability rates as prompt normativity weakens, with the most pronounced rise occurring from L2 to L3, matching the overall trends reported in the main results.
At the same time, the spread across models differs between the two groups: open-source models show larger between-model variability, especially under low-normativity prompts, whereas closed-source models appear more tightly clustered.
This divergence suggests that beyond prompt normativity, architectural and training choices contribute substantially to model-specific security behavior, and that scale alone is unlikely to explain the full variance.

\section{DISCUSSIONS}
Our experiments systematically show that a decrease in prompt normativity leads to a significant increase in security vulnerabilities. This section interprets these findings, discusses their broader implications, and acknowledges the study's limitations:

The core finding can be explained by a “path of least resistance” principle. When faced with a highly normative (L0) prompt, an LLM tends to adopt a professional “engineering” mindset, drawing upon safer practices from its knowledge base. However, when a prompt becomes vague or inconsistent (L2/L3), the LLM's task shifts from “implementation” to “interpretation.” To resolve this uncertainty, the model defaults to the most direct and simple implementation path. In security-critical tasks, this path of least resistance\citep{geirhos2020shortcut} is often the least secure, such as direct string concatenation instead of sanitization.

This has important implications for both developers and LLM providers. For developers, writing high-quality prompts is a key security practice. Treating an LLM as a partner that requires clear specifications, rather than a “black box”that can guess intent from vague requests, can significantly reduce risks. For LLM providers, future models should improve their “safe default” behavior under uncertainty, such as by asking clarifying questions or choosing more conservative and secure implementations when requirements are unclear.

%While our findings are robust, this study has limitations. Our experiments focused on single, independent Python functions; generalizability to more complex projects and other languages requires further validation. Future work could expand our benchmark to multi-module systems and explore how to fundamentally improve a model's robustness to vague prompts through fine-tuning.

While our findings are robust, this study has limitations. Our experiments focused on single, independent Python functions; generalizability to more complex projects and other languages requires further validation. Moreover, in real-world settings, low-quality prompts frequently arise within integrated development environments (IDEs) and agentic workflows, where incomplete, fragmented, or underspecified instructions may propagate across multiple steps and tools, potentially compounding their impact on the overall quality and security of generated code. These practical interactions remain underexplored and warrant systematic investigation. Future work could expand our benchmark to multi-module systems and explore how to fundamentally improve a model's robustness to vague prompts through fine-tuning.

\section{CONCLUSION}
The impact of everyday, non-malicious prompt quality on the security of code generated by Large Language Models has been a critical but understudied area.

%This paper provides the first systematic evidence of a strong causal relationship between prompt normativity and the security of LLM-generated code. Using our large-scale benchmark, CWE-BENCH-PYTHON, we clearly demonstrated that as prompts become less clear, complete, and logically consistent, the rate at which LLMs introduce security vulnerabilities significantly and consistently increases. We further showed that advanced prompting techniques like Chain-of-Thought and Self-Correction can effectively mitigate these risks.
This paper provides the first systematic evidence of a strong causal relationship between prompt normativity and the security of LLM-generated code. Using our large-scale benchmark, CWE-BENCH-PYTHON, we clearly demonstrated that as prompts become less clear, complete, and logically consistent, the rate at which LLMs introduce security vulnerabilities significantly and consistently increases. We further validated this conclusion across Java and C/C++ counterparts of the same scenarios, as well as under different model hyperparameter settings. We also showed that advanced prompting techniques like Chain-of-Thought and Self-Correction can effectively mitigate these risks.

Ultimately, our work shifts the focus of LLM code security research from solely examining the model's internal flaws or external malicious attacks to include the crucial dimension of human-AI interaction quality\citep{10.1145/3290605.3300233}. Our results make it clear that in the era of AI-assisted software development, crafting clear, complete, and logical requirements is not just key to functional efficiency—it is an essential security engineering practice.

\backmatter

\bibliography{sn-bibliography}% common bib file

\end{document}

%% file: attachments/tables/vulnerability_rates.tex
\begin{table*}[htbp]
\centering
\caption{Vulnerability Rates of Security Potential Validation Experiments with Optimization Methods Across Different Levels for Pillar-CWE Vulnerabilities.A “$\downarrow$” denotes a decrease relative to the corresponding Basic value. Colors encode rate bands: G (0–10\%), Y (10–20\%), OR (20–30\%), R ($\geq$30\%).}
\label{tab:vulnerability_rates}

\resizebox{\textwidth}{!}{
\begin{tabular}{lcccccccccccc}
\toprule
\textbf{Pillar-CWE} & \multicolumn{4}{c}{\textbf{Basic}} & \multicolumn{4}{c}{\textbf{cot}} & \multicolumn{4}{c}{\textbf{re-act}} \\
\cmidrule(lr){2-5} \cmidrule(lr){6-9} \cmidrule(lr){10-13}
 & L0(\%) & L1(\%) & L2(\%) & L3(\%) & L0(\%) & L1(\%) & L2(\%) & L3(\%) & L0(\%) & L1(\%) & L2(\%) & L3(\%) \\
\midrule
CWE-284 & \Y{13.59} & \Y{15.00} & \OR{27.81} & \R{49.84} &
\Y{13.41} $\downarrow$ & \Y{16.14} & \OR{20.91} $\downarrow$ & \R{43.41} $\downarrow$ &
\Y{14.64} & \Y{15.99} & \OR{23.95} $\downarrow$ & \R{42.41} $\downarrow$ \\
CWE-664 & \Y{18.38} & \OR{27.88} & \OR{24.38} & \R{34.62} &
\Y{19.87} & \OR{26.76} $\downarrow$ & \OR{22.64} $\downarrow$ & \R{30.55} $\downarrow$ &
\Y{18.75} & \OR{24.15} $\downarrow$ & \OR{20.78} $\downarrow$ & \OR{27.79} $\downarrow$ \\
CWE-682 & \G{4.79}  & \G{4.58}  & \G{4.79}  & \G{4.38}  &
\G{3.15} $\downarrow$ & \G{6.02} & \G{5.44} & \G{5.11} &
\G{4.59} $\downarrow$ & \G{3.21} $\downarrow$ & \G{4.14} $\downarrow$ & \G{2.35} $\downarrow$ \\
CWE-691 & \Y{14.84} & \Y{18.75} & \OR{22.03} & \R{36.41} &
\Y{15.02} & \Y{18.02} $\downarrow$ & \Y{19.82} $\downarrow$ & \R{33.33} $\downarrow$ &
\Y{14.24} $\downarrow$ & \Y{17.45} $\downarrow$ & \OR{21.00} $\downarrow$ & \OR{27.39} $\downarrow$ \\
CWE-693 & \R{31.87} & \R{32.38} & \R{35.50} & \R{38.00} &
\OR{24.88} $\downarrow$ & \OR{25.04} $\downarrow$ & \OR{27.56} $\downarrow$ & \OR{29.76} $\downarrow$ &
\Y{18.58} $\downarrow$ & \OR{20.93} $\downarrow$ & \Y{19.31} $\downarrow$ & \OR{27.99} $\downarrow$ \\
CWE-697 & \Y{11.25} & \Y{12.29} & \Y{10.00} & \Y{18.12} &
\Y{10.39} $\downarrow$ & \G{9.61} $\downarrow$ & \Y{11.17} & \Y{13.51} $\downarrow$ &
\Y{10.14} $\downarrow$ & \Y{12.72} & \Y{11.89} & \OR{20.10} \\
CWE-703 & \OR{23.12} & \OR{26.88} & \R{32.66} & \R{58.59} &
\OR{27.49} & \OR{27.61} & \R{33.97} & \R{49.79} $\downarrow$ &
\OR{28.57} & \OR{27.36} & \R{32.61} $\downarrow$ & \R{49.44} $\downarrow$ \\
CWE-707 & \R{31.25} & \R{32.62} & \OR{27.12} & \R{30.75} &
\Y{18.61} $\downarrow$ & \Y{18.96} $\downarrow$ & \Y{19.65} $\downarrow$ & \OR{25.57} $\downarrow$ &
\Y{14.59} $\downarrow$ & \Y{13.91} $\downarrow$ & \Y{13.74} $\downarrow$ & \Y{19.79} $\downarrow$ \\
\bottomrule
\end{tabular}}
\end{table*}